\documentclass[aps,amssymb,superscriptaddress,notitlepage,singlecolumn,longbibliography]{revtex4-1}

\makeatletter
\renewcommand*{\p@subsubsection}{}
\makeatother

\usepackage[letterpaper,margin=0.74in,footskip=0.25in]{geometry}

\usepackage{bm,amsmath,amssymb,
}
\usepackage{amsmath}
\usepackage{graphicx}
\usepackage{mathtools}
\usepackage{mathtools}
\usepackage[usenames,dvipsnames]{xcolor}
\usepackage[most]{tcolorbox}
\usepackage{color}
\usepackage{float}
\usepackage{empheq}
\usepackage{epstopdf}
\usepackage{epstopdf}
\usepackage{amssymb}
\usepackage{wrapfig}
\usepackage{picture}
\usepackage{mwe,tikz}
\usepackage[linktoc=all]{hyperref}
\usepackage{transparent}
\usepackage{textcomp}
\usepackage{enumerate}
\usepackage{overpic}
\usepackage[bold=1]{xfakebold}

\hypersetup{
	colorlinks,
	citecolor=blue,      
	filecolor=red,
	linkcolor=blue,
	urlcolor=blue,
	hyperfigures
}

\makeatletter
\newcommand{\verbatimfont}[1]{\def\verbatim@font{#1}}%
\makeatother

\newcommand{\uvc}[1]{\hat{\bm{\mathrm #1}}} 

\newcommand{\bff}{{\bf f}}
\newcommand{\bu}{{\bf u}}

\newcommand{\bx}{{\bf x}}

\newcommand{\md}{\mathrm d}

\newcommand{\mf}{\mathrm f}
\newcommand{\mh}{\mathrm h}

\begin{document}
	
	\title{A chemomechanical model of sperm locomotion reveals two modes of swimming}
	
	\author{Chenji Li} 
	\affiliation{Department of Mechanical and Aerospace Engineering, University of California San Diego, La Jolla, CA 92093, USA}
	\author{Brato Chakrabarti}
	\affiliation{Center for Computational Biology, Flatiron Institute, New York, NY 10010}
	\author{Pedro Castilla}
	\affiliation{Department of Mechanical and Aerospace Engineering, University of California San Diego, La Jolla, CA 92093, USA}
	\author{Achal Mahajan}
	\affiliation{Department of Mechanical and Aerospace Engineering, University of California San Diego, La Jolla, CA 92093, USA}
	\author{David Saintillan}
	\email{dstn@ucsd.edu}
	\affiliation{Department of Mechanical and Aerospace Engineering, University of California San Diego, La Jolla, CA 92093, USA}
	\date{\today}

	\begin{abstract}
		The propulsion of mammalian spermatozoa relies on the spontaneous periodic oscillation of their flagella. These oscillations are driven internally by the coordinated action of ATP-powered dynein motors that exert sliding forces between microtubule doublets, resulting in bending waves that propagate along the flagellum and enable locomotion. We present {an integrated} chemomechanical model of a freely swimming spermatozoon that uses a sliding-control model of the axoneme capturing the {two-way feedback between} motor kinetics {and} elastic deformations while accounting for {detailed fluid mechanics around the moving cell. We develop a robust computational framework that solves a boundary integral equation for the passive sperm head alongside the slender-body equation for the deforming flagellum described as a geometrically nonlinear internally actuated Euler-Bernoulli beam, and captures full hydrodynamic interactions.} Nonlinear simulations are shown to produce {spontaneous oscillations} with realistic beating patterns and trajectories, which we analyze as a function of sperm number and motor activity. Our {results indicate} that the swimming velocity does not vary monotonically with dynein activity, but instead displays two maxima corresponding to distinct modes of swimming, each characterized by qualitatively different waveforms and trajectories. {Our model also provides an estimate for the efficiency of swimming, which peaks at low sperm number.}
	\end{abstract}
	\maketitle

	\section{Introduction}
	
	The world at low Reynolds number comprises of a large variety of swimming microorganisms \cite{purcell1977life}. Examples range from spermatozoa that navigate through the female reproductive tract to fuse with the ovum, to ciliated unicellular organisms like \textit{Paramecium} commonly found in ponds, to bacteria found in guts to algae in the oceans \cite{lauga2020fluid}. These microorganisms rely on various mechanisms to break the time reversibility of Stokes flow in order to propel themselves in the suspending fluid \cite{lauga2009hydrodynamics}. While bacteria like \textit{Escherichia coli} use the rotation of their helical flagellar bundle for propulsion, eukaryotes like sperm cells rely on the propagation of bending waves along their flagella. Even though the nomenclature of flagellum is used for both prokaryotes and eukaryotes, their structure and origin are distinctly different. Eukaryotic flagella (or cilia) are thin hair-like cellular projections with an internal core known as the \textit{axoneme} that has been preserved during the course of evolution \cite{alberts2015essential}. The axoneme has a circular cross-section and is roughly $200 \ \mathrm{nm}$ in diameter with 9 pairs of microtubule  doublets arranged uniformly along its periphery. The doublets are connected with each other through a spring-like protein structure called nexin that extends along the entire length of the axoneme. Thousands of dynein molecular motors act between the microtubule doublets and generate internal sliding or shear forces in the presence of ATP. Due to structural constraints, the sliding forces are converted to internal bending moments that deform the flagellar backbone \cite{brokaw1972computer,riedel2007molecular}. Through a highly coordinated binding and unbinding, the molecular motors conspire to produce bending waves along the flagellum that help in the propulsion of spermatozoa \cite{chakrabarti2019spontaneous}. There have been several modeling efforts with varying levels of detail and complexity aimed at elucidating the biophysical processes that give rise to these spontaneous oscillations in isolated, and fixed filaments \cite{brokaw1972computer,brokaw1972flagellar,brokaw1999computer,brokaw2002computer,brokaw2005computer,brokaw2014computer,lindemann1994geometric,lindemann1994model,lindemann1996functional,holcomb1999flagellar,lindemann2002geometric,bayly2014equations,oriola2017nonlinear,chakrabarti2019hydrodynamic}. While the basic mechanisms giving rise to spontaneous  deformations are now well known, the detailed relationship between internal dynein actuation, elastohydrodynamics of the flagellum, non-local hydrodynamic interactions, and emergent waveforms and motility characteristics remains poorly understood. In this work, we present a biophysical model of sperm locomotion that integrates details of internal elasticity and hydrodynamic interactions with a  chemomechanical feedback loop for dynein activity within an idealized geometry. The model is applied to elucidate the relationship between internal actuation and the resulting beating patterns,   and demonstrates the key role of dynein activity in controlling the gait and overall motility of the spermatozoon. 
	
	The hydrodynamics of swimming sperm has been widely studied, going back to the classical work of G.I. Taylor on swimming sheets \cite{taylor1951analysis}. This has been followed by a series of mathematical analyses of flagellar propulsion \cite{taylor1952analysis,hancock1953self,gray1955propulsion,dresdner1980propulsion}, and hydrodynamic simulations \cite{higdon1979hydrodynamic,phan1987boundary,ramia1993role,fauci1995sperm}. Recent hydrodynamic studies
	relevant to sperm motility have addressed the role of surfaces in sperm accumulation \cite{elgeti2010hydrodynamics,smith2009human}, viscoelasticity of the medium \cite{teran2010viscoelastic,lauga2007propulsion}, and geometry of the head \cite{gillies2009hydrodynamic,gadelha2010nonlinear}. Almost all \cite{dillon2003mathematical} of such mathematical models coarse-grain the internal mechanics of the axoneme by prescribing the kinematics of the flagellum. 
	
	However, it is known that a variety of chemical cues related to calcium (Ca$^{2+}$) signaling along the axoneme regulate the flagellar beating. Such signaling pathways are responsible for motility \cite{olson2011coupling}, hyperactivation \cite{ho2001hyperactivation}, and the reversal of wave-propagation direction along the flagellum \cite{olson2010model}.  {As a first step towards understanding such biophysical phenomena, one needs to construct a model that incorporates the necessary chemomechanical feedback loops giving rise to sustained flagellar beating, coupled to all the relevant hydrodynamic interactions}.

	We address this in this paper by building on our previous work on active filaments used to model spontaneous oscillations of {isolated and fixed} cilia and flagella \cite{chakrabarti2019spontaneous}. The proposed biophysical model of a swimming spermatozoon includes the following: (a) a {simplified} model for flagellar beating that accounts for an idealized axonemal structure, internal elasticity, and dynein activity and kinetics, and (b) detailed non-local hydrodynamic interactions between the head and the flagellum. The paper is organized as follows. First, in Sec.~\ref{sec:model} we provide a brief description of the active filament model, the necessary boundary conditions, and outline the numerical method. We then discuss a linear stability analysis in Sec.~\ref{sec:result} followed by the analysis of various beating patterns and their properties far from equilibrium. By characterizing the swimming trajectories and emergent waveforms, we reveal how internal activity affects the motility and gives rise to two distinct modes of swimming. Using an energy budget, we then highlight the efficiency of the model spermatozoon. We finally discuss the features of both instantaneous  and time-averaged flow fields. We summarize and conclude in Sec.~\ref{sec:concl}.

	\section{Spermatozoon model}\label{sec:model}
	
	\subsection{Equations of motion}
	
\begin{figure}[t]
		\centering
		\includegraphics[width=0.9\linewidth]{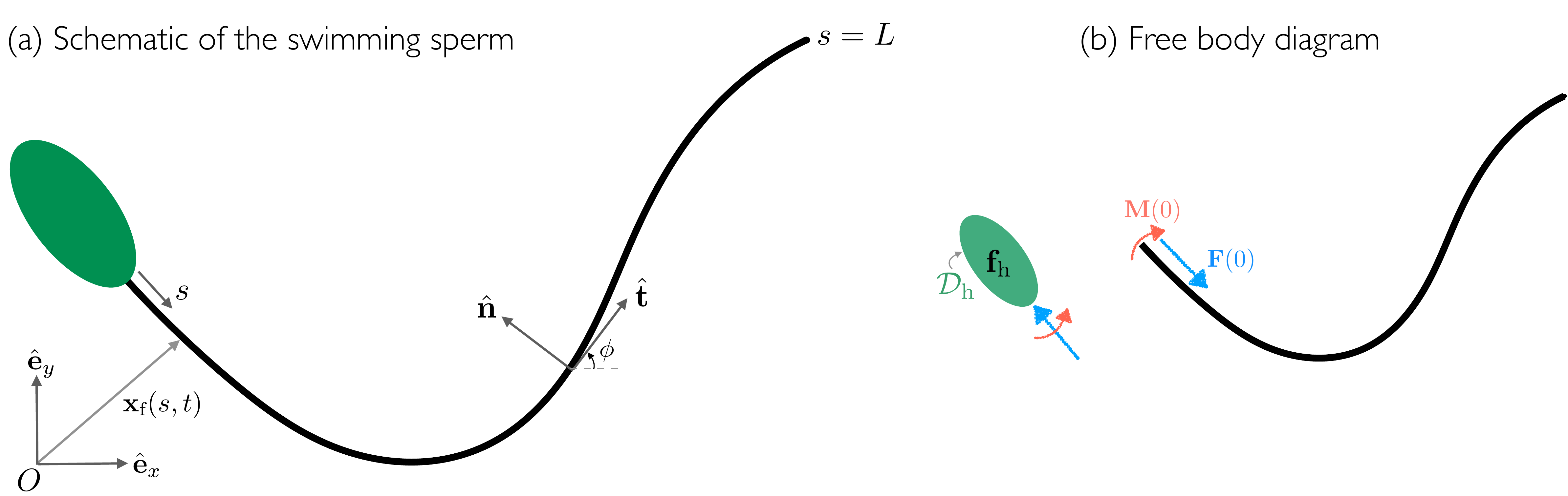}
		\caption{(a) Schematic of a swimming spermatozoon. In our model, a spheroidal head (shown in green) is clamped to a flexible flagellum described as a planar spacecurve with centerline $\bx_\text{f}$(s,t). (b) Free body diagram illustrating the balance of forces and moments at the head-flagellum junction.} \vspace{-0.2cm}
		\label{fig:schematic}
	\end{figure}

	A mature human sperm head is $5-6\, \mu$m long and $3\, \mu$m wide \cite{gaffney2011mammalian}. We choose to model it as a rigid spheroid. The flagellum of a human sperm cell has length $L \sim 30-50\,  \mu$m and cross-sectional diameter $a \sim 200$ nm. We model this using an active filament model \cite{chakrabarti2019spontaneous} that incorporates the necessary structural details of the axoneme, and accounts for various biophysical active processes that drive spontaneous oscillations. The active filament model approximates the 3D axoneme by its 2D projection. As a result, the beating of the flagellum in our model is entirely planar. As depicted in Fig.~\ref{fig:schematic}, the spheroidal head is clamped to the flagellum and is immersed in a 3D infinite fluid bath. 
	
	We parametrize the centerline of the active filament by its arc-length $s \in [0,L]$ and identify any point on it by the Lagrangian marker $\bx_\mathrm{f}(s,t)$ in a fixed reference frame. For an inextensible filament, we then have 
	\begin{equation}
		\mathbf{x}_\mathrm{f} (s,t) = \mathbf{x}_\mathrm{f} (0,t)+\int_{0}^{s}\hat{\mathbf{t}}(s',t) \ \md s',
	\end{equation}
	where $\uvc{t} = \cos \phi \ \uvc{e}_x +  \sin \phi \ \uvc{e}_y$ is the tangent to the centerline and $\phi(s,t)$ is the tangent angle as depicted in Fig.~\ref{fig:schematic}. We also define the associated unit normal along the centerline $\uvc{n} = -\sin \phi \ \uvc{e}_x + \cos \phi \ \uvc{e}_y$. The velocity at any point along the filament is then given by 
	\begin{equation}
		\mathbf{v} (s,t) \equiv \dot{\mathbf{x}}_\mathrm{f}(s,t) = \mathbf{v} (0,t)+\int_{0}^{s} \Dot{\phi}(s',t) \hat{\mathbf{n}}(s',t) \ \md s'.
		\label{eq:velcomp}
	\end{equation}
	
	Force and torque balance for a planar elastic rod in the overdamped limit yields \cite{chakrabarti2019spontaneous}
	\begin{align}
		\bff_{\text{vis}} + \partial_s \mathbf{F} &=0 \label{eq:forcebal}, \\
		\partial_s M + N &=0 \label{eq:torbal},
	\end{align}
	where $\bff_{\text{vis}}$ is the viscous force per unit length exerted by the fluid on the filament, $\mathbf{F} = T \uvc{t} + N \uvc{n}$ is the contact force, and $\mathbf{M} = M \uvc{e}_z$ is the contact moment \cite{Antman:1250280} in the active filament. Since the flagellum is a slender filament ($\epsilon = a/L \ll 1$), we model its hydrodynamics using non-local slender body theory (SBT) \cite{KR1976,tornberg2004simulating}, which relates viscous forces to the centerline velocity as 
	\begin{equation}\label{eq:SBT}
		8\pi\nu \left(\mathbf{v} - \mathcal{H}[\bff_\mathrm{h}]\right) = -\mathcal{M} [\mathbf{f}_{\text{vis}}] \equiv \mathcal{M}[\bff_\mf].
	\end{equation}
	Here, $\nu$ is the fluid viscosity. The term $\mathcal{H}[\bff_\mh]$, where $\bff_\mh$ is the hydrodynamic traction on the sperm head, denotes the disturbance velocity due to the motion of the head and is obtained as a single-layer boundary integral equation \cite{pozrikidis1992}:
	\begin{equation}
		\mathcal{H} [\mathbf{f}_\mh](s) = \iint _{\mathcal{D}_\mh} \textbf{G}(\mathbf{x}_\mf(s),\mathbf{x}_\mh) \cdot \mathbf{f}_\mh(\mathbf{x}_\mh) \ \md S(\bx_\mh),
	\end{equation}
	where $\mathbf{G}(\bx,\bx')$ is the 3D free-space Green's function for Stokes flow, and $\mathcal{D}_\mh$ denotes the 2D surface of the head. The right-hand side in Eq.~(\ref{eq:SBT}) involves the force per unit length exerted by the flagellum on the fluid $\bff_\mf \equiv \partial_s \mathbf{F}=-\bff_{\text{vis}}$ through a mobility operator  $\mathcal{M}$ with two contributions: $\mathcal{M} = \mathcal{L} + \mathcal{K}$ \cite{KR1976}.
	The local part $\mathcal{L}[\mathbf{f}_\mf]$ accounts for drag anisotropy along the flagellum and is given by
	\begin{equation}
		\mathcal{L}[\bff_\mf](s)=\left[\frac{1}{\xi_\perp}\hat{\bold{n}}(s)\hat{\bold{n}}(s)+\frac{1}{\xi_\parallel}\hat{\bold{t}}(s)\hat{\bold{t}}(s)\right]\cdot \bff_\mf(s),
	\end{equation}
	where $\xi_\perp= (2-c)^{-1}$ and $\xi_\parallel = -(2c)^{-1}$ are resistance coefficients in the normal and tangential directions, and $c = \ln(\epsilon^2 \mathrm{e})< 0$. Non-local hydrodynamic interactions between distant flagellar sections are captured by $\mathcal{K} [\mathbf{f}_\mathrm{f}]$ defined as
	\begin{equation}
		\mathcal{K}[\bff_\mf](s)= 
		\int_0^L \Biggl[ 
		\frac{\bold{I}+\hat{\bold{R}}(s,s')\hat{\bold{R}}(s,s')}{|\bold{R}(s,s')|}
		\cdot \bff_\mf(s') -
		\frac{\bold{I}+\hat{\bold{t}}(s)\hat{\bold{t}}(s)}{|s-s'|}
		\cdot \bff_\mf(s)
		\Biggr]
		\md s',
	\end{equation}
	where $\mathbf{R}(s,s') = \mathbf{x}_\mf(s) - \mathbf{x}_\mf(s')$ and $\hat{\mathbf{R}} = \mathbf{R}/|\mathbf{R}|$. 
	
	\subsection{Active filament model}
	Here, we provide a concise overview of the active filament model for the flagellum, which directly follows our past work on clamped filaments \cite{chakrabarti2019spontaneous} as well as a prior model by Oriola \textit{et al.} \cite{oriola2017nonlinear}. {These build on an earlier model by Riedel-Kruse \textit{et al.}~\cite{riedel2007molecular} and on seminal work by Brokaw \cite{brokaw1972computer,brokaw1972flagellar}}. The interested reader is pointed to these references for further details. We idealize the 3D axoneme by its planar projection in the plane of motion, described as an elastic structure of width $a$, length $L$, and centerline $\bx_\mf(s)$. In this projection, microtubules from the opposite sides of the axoneme are represented by two polar filaments $\bx_{\pm} = \bx_\mf \pm a \uvc{n}/2$ clamped at the base at $s=0$ and connected to one another by passive nexin crosslinkers as well as dynein motors, which exert shear forces $\pm  f_m(s) \uvc{t}(s)$ per unit length. These forces result in a
	sliding displacement $\Delta(s,t)= a \left(\phi(s,t)-\phi(0,t)\right)$ between the 
	two filaments. The sliding force density can be expressed as
	\begin{equation}
		f_{m}(s, t)= \rho\left(n_{+} F_{+}+n_{-} F_{-}\right)-K \Delta(s, t),
	\end{equation}
	where $\rho$ is the line density of motors, $n_\pm(s,t)$ is the fraction of bound motors on $\bx_\pm$, $F_\pm$ is the force exerted by an individual dynein, and $K$ is the stiffness of nexin links modeled as linear springs. The force exerted by the motors follows a linear force- velocity relation $F_\pm = \pm f_0(1 \mp \dot{\Delta}(s, t)/v_0)$, where $f_0$ is the stall force of dynein, $\dot{\Delta}$ is the sliding velocity, and $v_0$ is a characteristic velocity scale. The inability of the microtubules to freely slide apart means that the sliding forces give rise to an active bending moment 
	\begin{equation}
		\mathbf{M}(s, t)= \uvc{e}_z \left[B \phi_{s}(s,t) - a \int_s^L f_m(s',t) \ \md s'\right],
	\end{equation}
	where $B$ is the bending rigidity of the flagellum. Moment balance in the out-of-plane direction from equation~\eqref{eq:torbal} yields
	\begin{equation}
		B \phi_{ss} + a f_m + N = 0.
	\end{equation}
	Here and in the following, subscript $s$ denotes differentiation with respect to arc-length. The attachment and detachment of dynein motors between the two filaments follow first-order kinetics: $\dot{n}_\pm = \pi_\pm - \epsilon_\pm + D \partial_s^2 n_\pm$. The attachment rate $\pi_\pm = \pi_0 (1-n_\pm)$ is proportional to the unbound motor population with a characteristic rate constant $\pi_0$. We use a force dependent detachment rate as $\epsilon_\pm = \epsilon_0 n_{\pm} \exp(|F_{\pm}|/f_c)$. Here, $\epsilon_0$ is the characteristic rate of detachment and $f_c$ is the force scale above which the motors detach exponentially fast. {Note that more sophisticated models may include a dependence of the detachment rate on local curvature \cite{sartori2019effect,chakrabarti2019spontaneous}, though we neglect this feature here.} The last term {in the kinetic equation} accounts for diffusion of motors along the filament backbone with diffusivity $D$. 
	
	\subsection{Dimensionless governing equations}
	
	We  non-dimensionalize the arc-length by $L$, and the sliding displacement by the axoneme diameter $a$. The characteristic time scale for the problem is the motor correlation time $\tau_0 = 1/(\epsilon_0 + \pi_0)$. The density of the sliding force between the filaments is scaled by $\rho f_0$, and the elastic force in the flagellum is scaled by $B/L^2$. This results in four dimensionless groups:
	\begin{itemize}
		\item $\mathrm{Sp} = L(8\pi\nu\xi_\perp/B\tau_0)^{1/4}$ is the so-called sperm number, comparing the time scale of bending relaxation to the motor correlation time. Larger sperm number corresponds to a more flexible flagellum.
		\item $\mu_a = a\rho f_0L^2/B$ compares the active force to the passive bending force and is a measure of the activity of the flagellum. 
		\item $\mu = K a^2L^2/B$ is the ratio between resistance from the nexin links and bending elasticity.
		\item $\zeta = a/v_0\tau_0$ compares the diameter of the flagellum to the characteristic displacement due to motor activity.
	\end{itemize}
	We also define the duty ratio $\eta = \pi_0/(\epsilon_0 + \pi_0)$, and $\Bar{f} = f_0/f_c$. The dimensionless governing equations are now given as:
	\begin{equation}
		T_{ss}-N\phi_{ss}-
		\left(1+\frac{\xi_\parallel}{\xi_\perp}\right) N_s\phi_s-\frac{\xi_\parallel}{\xi_\perp} T\phi_s^2
		= \xi_\parallel \left(\phi_s u_n^d - \partial_s u_t^d\right),
		\label{eq:tanbal}
	\end{equation}
	\begin{equation}
		N_{ss}+T\phi_{ss}+ 
		\left(1+\frac{\xi_\perp}{\xi_\parallel}\right)T_s\phi_s-\frac{\xi_\perp}{\xi_\parallel} N\phi_s^2 \\
		= \mathrm{Sp}^4\dot{\phi} - \xi_\perp \left(u_t^d \phi_s + \partial_s u_n^d \right),
		\label{eq:norbal}
	\end{equation}
	\begin{equation}
		\phi_{s s}+\mu_a f_m+N=0,
		\label{eq:mombal}
	\end{equation}
	\begin{equation}
		f_m = \Bar{n}-\zeta \Tilde{n} \dot{\Delta} - \frac{\mu}{\mu_a}\Delta,
		\label{eq:active}
	\end{equation}
	\begin{equation}
		\dot{n}_\pm = \eta (1-n_\pm) - (1-\eta) n_\pm \exp\left[\Bar{f}(1\mp \zeta \dot{\Delta})\right] + D \partial_s^2 n_\pm.
		\label{eq:dynein}
	\end{equation}
	Equations~\eqref{eq:tanbal} and \eqref{eq:norbal} represent the force balance in the tangential and normal directions and were obtained by differentiating the slender-body equation (\ref{eq:SBT}) with respect to arc-length, and equation~\eqref{eq:mombal} is the dimensionless moment balance. We have introduced the following definitions: $\bar{n} = n_+ - n_-$ and $\tilde{n} = n_+ + n_-$. In equations~\eqref{eq:tanbal} and \eqref{eq:norbal}, $u^d_t$ and $u^d_n$ denote the tangential and normal components of the disturbance velocity due to to tractions on the both head and flagellum, which captures non-local hydrodynamic interactions:
	\begin{align}
		\mathbf{u}^d(s)=\mathcal{K}[\mathbf{f}_\mf] (s) + \mathcal{H}[\mathbf{f}_\mh] (s) .  \label{eq:veldist}
	\end{align}\vspace{-0.25cm}

	\subsection{Boundary conditions}
	The head of the spermatozoon has no motility and is pushed, dragged, and rotated by the force and torque coming from the flagellum. The free body diagram in Fig.~\ref{fig:schematic}(b) illustrates the forces acting on each component of the sperm model.
	The boundary condition {at $s = 0$} is given by the force and the moment balance equations
	\begin{align}
		\mathbf{F}(0) - \iint _{\mathcal{D}_\mh} \mathbf{f}_\mh (\bx_\mh) \md S(\bx_\mh) &= \mathbf{0}
		\label{eq:BC1}, \\
		\mathbf{x}_\mf(0) \times \mathbf{F}(0) + \mathbf{M}(0) - \iint _{\mathcal{D}_\mh} \bx_\mh \times \mathbf{f}_\mh (\bx_\mh) \md S(\bx_\mh) &= \mathbf{0}
		\label{eq:BC2}.
	\end{align}
	The distal end of the flagellum is force and moment-free, which implies 
	\begin{align}\label{eq:FM1}
		\mathbf{F}(1) = 0, \qquad
		\mathbf{M}(1) = 0.
	\end{align}
	In order to compute the unknown hydrodynamic traction $\bff_\mh$, we make use of the no-slip boundary condition on the sperm head, which reads:
	
		\begin{align}
			\int_0^1 \!\textbf{G}(\mathbf{x}_\mh, \mathbf{x}_\mf(s))\cdot 
			\mathbf{f}_\mf(\mathbf{x}_\mf(s)) \md s + \iint _{\mathcal{D}_h}\!\!\! \textbf{G}(\mathbf{x}_\mh, \mathbf{x}'_\mh) \cdot\mathbf{f}_\mh(\mathbf{x}'_\mh) \md S(\bx'_\mh)   
			= \frac{\mathrm{Sp}^4}{\xi_\perp}\left[\mathbf{v}(0) + \Dot{\phi} (0) \uvc{e}_z \times \left(\mathbf{x}_\mh-\mathbf{x}_\mf(0)\right)\right], \label{eq:noslip}
		\end{align}
where the right-hand side captures the rigid body motion of the head. Finally, we use a no-flux boundary condition for the motor population, which translates to $\partial_s n_\pm = 0$ at $s = 0,1$.

\subsection{Numerical methods}
{The system of governing equations involves equations (\ref{eq:tanbal})--(\ref{eq:veldist}) for the flagellum elastohydrodynamics and internal motor kinetics and actuation, coupled to the boundary integral equation (\ref{eq:noslip}) on the surface of the sperm head. The coupling occurs through the clamped boundary conditions at the head-flagellum junction as well as through the net force and torque balance on the assembly as given in equations (\ref{eq:BC1})--(\ref{eq:BC2}). To numerically solve this system we first combine these governing equations to yield a linear system for the unknown contact forces $\{T,N\}$  \cite{chakrabarti2019hydrodynamic}. The combined equations are given as:
	\begin{align}
		T_{ss}-N\phi_{ss}-
		\left(1+\frac{\xi_\parallel}{\xi_\perp}\right) N_s\phi_s-\frac{\xi_\parallel}{\xi_\perp} T\phi_s^2
		& = \xi_\parallel \left(\phi_s u_n^d - \partial_s u_t^d\right)\Big|^{t-\Delta t} \label{eq:ten}
	\end{align}\vspace{-0.7cm}
	\begin{equation}
		\begin{split}
			&\mu_a \zeta \Tilde{n} \left[
			N_{s s}+T\phi_{s s}+\left(1+\frac{\xi_\perp}{\xi_\parallel}\right) T_s\phi_s-\frac{\xi_\perp}{\xi_\parallel} N\phi_s^2
			\right] - \mathrm{Sp}^4 N - \mathrm{Sp}^4 \mu_a \zeta \Tilde{n} \,{\Dot{\phi}(0)}
			\\
			& \qquad\qquad\qquad = \mathrm{Sp}^4 \left[
			\phi_{s s} + \mu_a \Bar{n} -\mu \left(\phi-\phi(0)\right)
			\right]
			-\xi_\perp \mu_a \zeta \Tilde{n}
			\left(u_t^d \phi_s + \partial_s u_n^d\right)\Big|^{t-\Delta t}.
		\end{split}
		\label{eq:normcomb}
	\end{equation}
	For a given flagellar shape $\phi$ and given bound motor distributions $n_{\pm}$, the unknowns in Eqs.~(\ref{eq:ten})--(\ref{eq:normcomb}) are the contact forces $\{T,N\}$, the disturbance velocity $\mathbf{u}^d$, as well as the angular velocity ${\dot{\phi}(0)}$ at the base, whose treatment we explain further below. In our simulations, the disturbance velocity $\mathbf{u}^d$ is treated explicitly and determined from the forces computed from the previous time step at $t-\Delta t$. At $t=0$, they are not included in the computation. An iteration scheme outlined in \cite{chakrabarti2019hydrodynamic} can be incorporated to improve the accuracy, but our numerical investigations suggest that it does not alter the present results. 
	
	The boundary conditions on the tangential and normal forces are given by:
	\begin{align}
		T(1) = N(1) &= 0 \label{eq:bc00}, \\ 
		\frac{\mathrm{Sp}^4}{\xi_\perp} {v_n(0)} - \frac{1}{\xi_\perp}(N_s + T\phi_s) &= u_n^d\big|^{t-\Delta t} \label{eq:bc01}, \\
		\frac{\mathrm{Sp}^4}{\xi_\perp} {v_t(0)} - \frac{1}{\xi_\parallel}(T_s - N\phi_s) &= u_t^d\big|^{t-\Delta t} \label{eq:bc02},
	\end{align}
	where the first equation simply specifies the force-free boundary condition at the distal end. The boundary conditions at $s=0$ reflect the local force balance and involve the tangential and normal velocity components ${v_t(0)}$ and ${v_n(0)}$ at the flagellum base, which are unknown and must match those of the head at that point.  
	
	Equations (\ref{eq:ten})--(\ref{eq:normcomb}), along with boundary conditions (\ref{eq:bc00})--(\ref{eq:bc02}), constitute a boundary value problem for the contact forces $\{T,N\}$. We discretize the flagellum along its arc length $s$ with  $N_\text{f}$ points uniformly distributed along the flagellum such that $\Delta s = 1/(N_\text{f}-1)$, and the spatial derivatives along the flagellum are then approximated with  a second-order-accurate finite difference scheme. This results in an algebraic systems of equations for the values of $\{T,N\}$ at the grid points. We note, however, that the three variables   ${\{v_t(0),v_n(0),\dot{\phi}(0)\}}$ remain unknown: these are the linear and angular velocities at the base of the flagellum, which couple the motion of the flagellum to that of the head and must be determined self-consistently along with the solution for the tension and normal forces. To this end, the dynamics of the spermatozoon head must be analyzed, as we explain next.

	Equations (\ref{eq:BC1})--(\ref{eq:FM1}) constitute a linear system for the traction field $\mathbf{f}_\mathrm{h}$ on the head surface, linear velocity $\mathbf{v}(0)$ of the head and angular velocity $\dot{\phi}(0)$; however, it also involves forces along the flagellum via the terms involving $\mathbf{f}_\mathrm{f}$, $\mathbf{F}(0)$ and $\mathbf{M}(0)$, and is thus fully coupled with the governing equations for  the flagellum motion. Here, we outline a method that allows us to eliminate the traction field $\mathbf{f}_\mathrm{h}$ and recast Eqs. (\ref{eq:BC1})--(\ref{eq:FM1}) into a linear system for ${\{v_t(0),v_n(0),\dot{\phi}(0)\}}$ that can be combined with the governing equations for $\{T,N\}$. 
	
	\begin{figure}[t]
		\centering
		\includegraphics[scale=0.25]{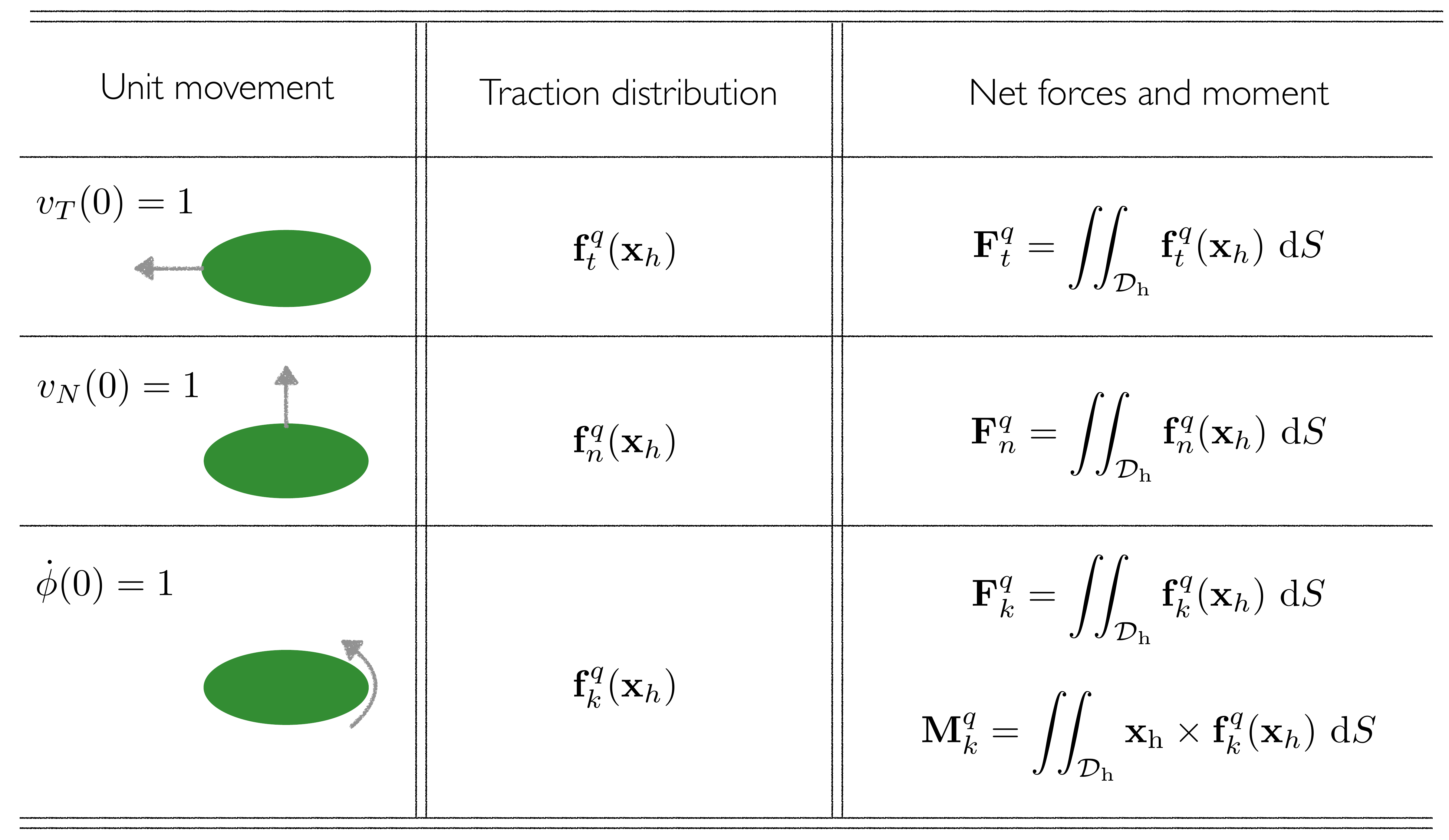}
		\caption{Illustration of the unit translations and rotation of the spermatozoon head, and their corresponding traction fields and net forces and moment.}
		\label{tab:response}
	\end{figure}
	
	In order to approximate the surface integrals over the head surface $\mathcal{D}_\mathrm{h}$ in Eqs. (\ref{eq:BC1})--(\ref{eq:FM1}), we use the boundary element method  \cite{pozrikidis2002practical} and triangulate the ellipsoidal head surface with $N_\text{h}$ triangles. The integrals are then computed by Gaussian quadrature on each element. In discrete form, Eq.~(\ref{eq:noslip}) can be written as
	\begin{equation}
		\mathbf{A}_{\mh \mh} \cdot {\textbf{f}_\mh} + \mathbf{A}_{\mh \mf} \cdot {\textbf{f}_\mf} = \frac{\mathrm{Sp}^4}{\xi_\perp} \left[{\textbf{v}(0)} + \Dot{\phi} (0) {\hat{\textbf{k}}\times \left(\mathbf{x}_\mh-\mathbf{x}_\mf(0)\right)}
		\right], \label{eq:discrBI}
	\end{equation}
	where $\mathbf{A}_{\mh \mh}$ is a $3 N_\mh \times 3 N_\mh$ matrix and $\mathbf{A}_{\mh \mf}$ is a $3 N_\mh \times 3 N_\mf$ matrix that encodes the interaction between the elements on the head and the flagellum. While $\mathbf{A}_{\mh \mf}$ changes at every time step as the flagellum deforms, we note that $\mathbf{A}_{\mh \mh}$ is only a function of the mesh geometry on the head surface: therefore, it does not change with time although its component must be rotated as for a second-order tensor as the head orientation changes.  
	
	Inverting Eq.~(\ref{eq:discrBI}) and decomposing the filament base velocity into tangential and normal components yields
	\begin{equation}
		\bff_\mh = -\textbf{A}_{\mh \mh}^{-1} \cdot \textbf{A}_{\mh \mf} \cdot {\textbf{f}_\mf} + v_t(0)\mathbf{f}_T^q + v_n(0)\mathbf{f}_N^q + \Dot{\phi}(0)\mathbf{f}_K^q.
	\end{equation}
	where 
	\begin{equation}
		\mathbf{f}_t^q=\textbf{A}_{\mh \mh}^{-1}\cdot \hat{\mathbf{t}}, \ \ \  \mathbf{f}_n^q=\textbf{A}_{\mh \mh}^{-1}\cdot \hat{\mathbf{n}}, \ \ \ \mathbf{f}_k^q=\textbf{A}_{\mh \mh}^{-1}\cdot\left[\hat{\textbf{k}}\times \left(\mathbf{x}_\mh-\mathbf{x}_\mf(0)\right)\right].  \label{eq:fheq}
	\end{equation}
	After inserting Eq.~(\ref{eq:fheq}) into Eqs.~(\ref{eq:BC1})--(\ref{eq:BC2}), we arrive at
	\begin{align}
		& v_t(0) \mathbf{F}_t^q + v_n(0) \mathbf{F}_n^q +  \Dot{\phi}(0) \mathbf{F}_k^q
		= \iint _{\mathcal{D}_h} \left(\textbf{A}_{\mh \mh}^{-1} \cdot \textbf{A}_{\mh \mf} \cdot {\textbf{f}_\mf}\right)\md S + \textbf{F}(0),  \label{eq:implicit1} \\
		& \Dot{\phi}(0) \mathbf{M}_k^q
		= \iint _{\mathcal{D}_h} \bx_\mh \times \left(\textbf{A}_{\mh \mh}^{-1} \cdot \textbf{A}_{\mh \mf} \cdot {\textbf{f}_\mf}\right)  \md S + \mathbf{x}_\mf(0) \times \textbf{F}(0) + \textbf{M}(0),
		\label{eq:implicit2}
	\end{align}
	where
	\begin{equation}
		\mathbf{F}^q_t=\iint_{\mathcal{D}_\mathrm{h}} \mathbf{f}_t^q(\mathbf{x}_\mathrm{h})\mathrm{d}S, \quad \mathbf{F}^q_n=\iint_{\mathcal{D}_\mathrm{h}} \mathbf{f}_n^q(\mathbf{x}_\mathrm{h})\mathrm{d}S, \quad \mathbf{F}^q_k=\iint_{\mathcal{D}_\mathrm{h}} \mathbf{f}_k^q(\mathbf{x}_\mathrm{h})\mathrm{d}S, 
	\end{equation}
	and
	\begin{equation}
		\mathbf{M}^q_k =\iint_{\mathcal{D}_\mathrm{h}} (\mathbf{x}_\mathrm{h}-\mathbf{x}_\mathrm{f}(0))\times \mathbf{f}_k^q(\mathbf{x}_\mathrm{h})\mathrm{d}S,
	\end{equation}
	are the forces and moment induced by unit translations and rotations of the head as depicted in Fig.~\ref{tab:response}. Note that these quantities only depend on the geometry of the head and its mesh: they can therefore be precomputed at the start of the simulation, as well as the matrix $\mathbf{A}_\mathrm{hh}$ and its inverse in a fixed reference frame. Projecting Eq.~(\ref{eq:implicit1}) on the tangential and normal directions and projecting Eq.~(\ref{eq:implicit2}) along $\hat{\mathbf{k}}$ provides three linear equations relating ${\{v_t(0),v_n(0),\dot{\phi}(0)\}}$ to the flagella variables $\mathbf{f}_\mathrm{f}$, $\mathbf{F}(0)$ and $\mathbf{M}(0)$, which are all linear functions of $\{T,N\}$. Combined with Eqs.~(\ref{eq:tanbal})--(\ref{eq:norbal}) and boundary conditions (\ref{eq:bc00})--(\ref{eq:bc02}), they provide a linear system that can inverted to solve for ${\{T, N,v_t(0),v_n(0),\dot{\phi}(0)\}}$ simultaneously at every instant.



\subsection{Parameter selection}
Following \cite{chakrabarti2019hydrodynamic}, we estimate model parameters from various experiments on cilia and flagella, and typical dimensional values are reported in table~\ref{tb:dimpar}. The corresponding dimensionless parameter ranges are given in table~\ref{tb:parameters}. 
The key dimensionless groups of our problem are the sperm number ($\mathrm{Sp}$) and the activity ($\mu_a$) number: we focus the following discussion on analyzing the emergence of spontaneous beating patterns and motility characteristics in terms of these two parameters. 

\begin{table*}[t]
	\centering
	\begin{tabular}{llll}
		\hline
		Parameter &Range &Dimension &Description \\
		\hline
		$L$ & $30-50$ & $\mu \mathrm{m}$ & Length of human sperm \cite{riedel2007molecular}\\
		$a$ & $200$ & $\mathrm{nm}$ & Effective diameter of axoneme \cite{gaffney2011mammalian}\\
		$B$ & $0.9 - 1.7 \times 10^{-21}$ & $\mathrm{N}\cdot\mathrm{m}^2$ & Range of bending rigidity of sea-urchin sperm and bull sperm \cite{riedel2007molecular, bayly2014equations}\\
		$K$ & $2 \times 10^{3}$& $\mathrm{N}\cdot\mathrm{m}^{-2}$ & Interdoublet elastic resistance measured for \textit{Chlamydomonas} \cite{oriola2017nonlinear}\\
		$\xi_\perp$ & $10^{-3} - 1$& $\mathrm{Pa}\cdot\mathrm{s}$ & Range of coefficient of normal drag in different viscous media \cite{oriola2017nonlinear,bayly2014equations}\\
		$f_0$ & $1 - 5$& $\mathrm{pN}$ & Stall force for motor dynamics \cite{oriola2017nonlinear}\\
		$f_c$ & $0.5 - 2.5$& $\mathrm{pN}$ & Characteristic unbinding force of the motors \cite{howard2001mechanics}\\
		$v_0$ & $5 - 7$& $\mu \mathrm{m}\cdot\mathrm{s}^{-1}$ & Motor walking speed at zero load \cite{howard2001mechanics}\\
		$\tau_0$ & $50$& $\mathrm{ms}$ & Correlation time of motor activity \cite{oriola2017nonlinear}\\
		$\rho$ & $10^3$& $\mu \mathrm{m}^{-1}$ & Mean number density of motors \cite{oriola2017nonlinear}\\
		\hline
	\end{tabular}
	\caption{Typical values of the dimensional parameters used in our simulations, as estimated from various experiments.}
	\vspace*{-8pt}\label{tb:dimpar}
\end{table*}

\begin{table}[t]
	\centering
	\begin{tabular}{ll}
		\hline
		Dimensionless number &Range \\
		\hline
		$\mathrm{Sp} = L(8\pi\nu\xi_\perp/B\tau_0)^{1/4}$ & $1 - 12$ \\
		$\mu_a = a\rho f_0L^2/B$ & $2 - 15 \times 10^{3}$\\
		$\bar{f} = f_0 / f_c$ & $2$ \\
		$\mu = K a^2L^2/B$ & $100$\\
		$\zeta = a/v_0\tau_0$ & $0.4$\\
		\hline
	\end{tabular}
	\caption{Dimensionless groups and their typical values.}
	\vspace*{-0pt}\label{tb:parameters}
\end{table}

\section{Results and discussion}\label{sec:result}

\subsection{Linear stability analysis \label{sec:stability}}
We first provide a brief discussion of the linear stability analysis that closely follows \cite{oriola2017nonlinear}. The base state of our problem is a straight undeformed filament. For simplicity, we assume a spherical head and neglect any hydrodynamic interactions in this section. We consider perturbations from this straight configuration of the form $\phi(s) = \varepsilon \Phi(s) \mathrm{e}^{\sigma t}$, where $\sigma$ is the growth rate and $\Phi(s)$ is the associated mode shape. The linearized system of governing equations yields an eigenvalue problem for the unknown growth rate $\sigma$.

The results from this analysis are summarized in Fig.~\ref{fig:linstab}. The dependence of the real and imaginary parts of $\sigma$ on activity $\mu_a$ for a fixed sperm number of $\mathrm{Sp}=8$ is shown in Fig.~\ref{fig:linstab}(a). At low levels of activity, $\mathrm{Re}[\sigma]\le 0$, indicating that the straight equilibrium configuration is linearly stable to perturbations, i.e. the sperm cell remains undeformed and does not swim. Upon increasing $\mu_a$, a Hopf bifurcation takes place, above which $\mathrm{Re}[\sigma]>0$ and $\mathrm{Im}[\sigma]\neq 0$, indicating the spontaneous time-periodic oscillation of the active filament and subsequent swimming of the sperm cell. As the level of activity keeps increasing, the growth rate $\mathrm{Re}[\sigma]$ also increases while the magnitude of $\mathrm{Im}[\sigma]$ decreases monotonically until it finally reaches zero, marking a second bifurcation above which the linear theory no longer predicts oscillations. This second bifurcation is also accompanied by an increase in the slope of $\mathrm{Re}[\sigma]$. The critical activity level for both bifurcations increases monotonically with $\mathrm{Sp}$ and is plotted in the ($\mathrm{Sp}$, $\mu_a$) parameter space in Fig.~\ref{fig:linstab}(b). Similar trends had been predicted in past studies  \cite{oriola2017nonlinear,chakrabarti2019spontaneous,man2020cilia} considering fixed filaments that were either clamped or hinged at the base: there, the second bifurcation in the case of clamped boundary conditions was shown to be associated with a reversal in the direction of wave propagation along the flagellum in the nonlinear regime, with retrograde (tip-to-base) propagation below the second bifurcation switching to anterograde (base-to-tip) propagation above \cite{chakrabarti2019spontaneous,man2020cilia}. As we will see in the nonlinear simulations of the next section, the behavior is different in the case of a freely swimming cell: spontaneous oscillations with base-to-tip propagation are indeed observed both below and above the second bifurcation, but display a qualitative change in the beating and swimming behavior across the transition. 

\begin{figure}[t]
	\centering\vspace{-0.0cm}
	\includegraphics[width=0.85\linewidth]{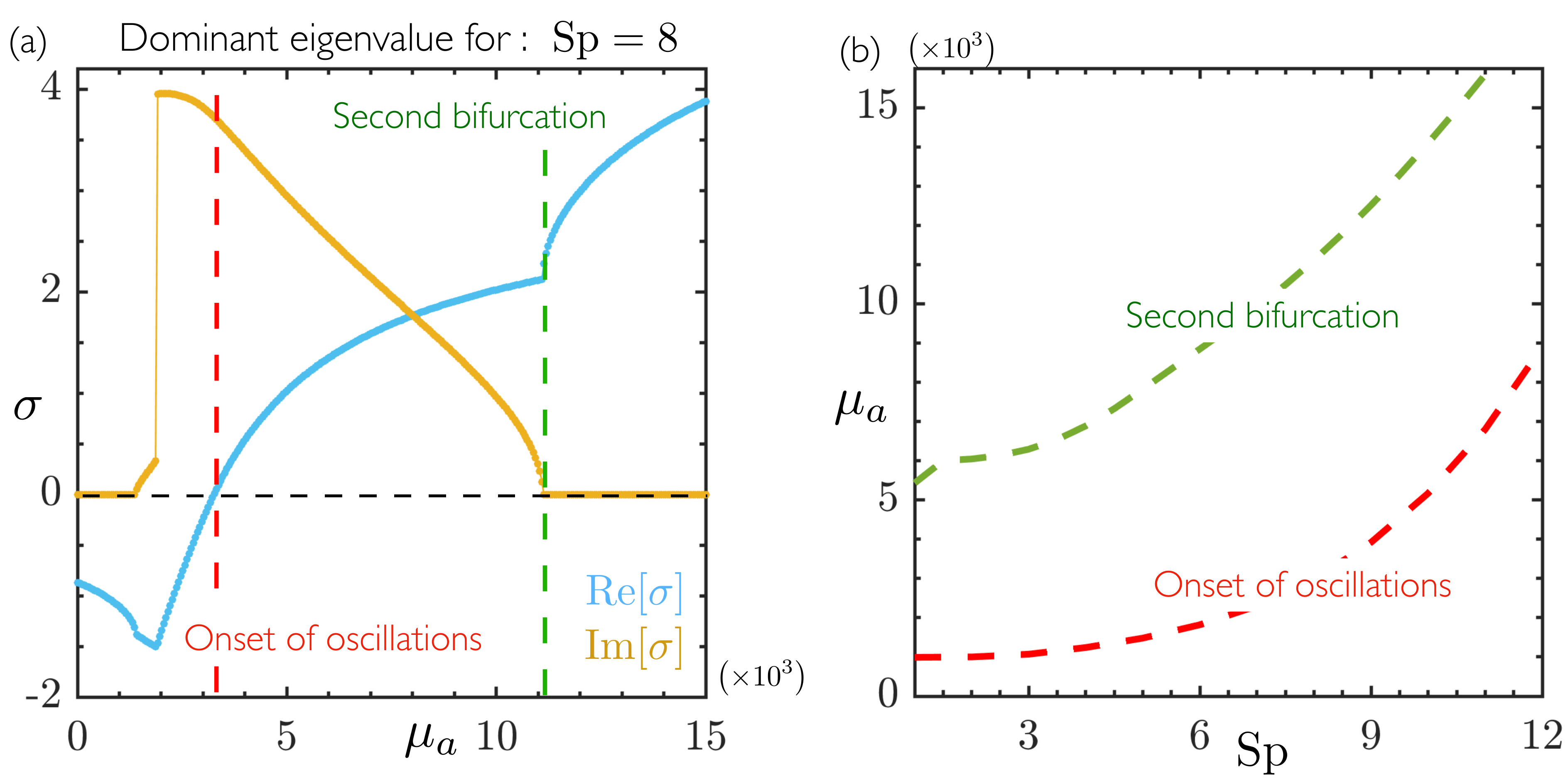}\vspace{-0.0cm}
	\caption{(a) Real and imaginary parts of the dominant eigenvalue $\sigma$ obtained in the linear stability as a function of activity parameter $\mu_a$ at a fixed sperm number of $\mathrm{Sp}=8$. The vertical dashed lines highlight the Hopf bifurcation marking the onset of spontaneous oscillations, and the second bifurcation above which $\mathrm{Im}[\sigma]=0$. (b) Phase diagram highlighting the thresholds for the Hopf bifurcation (red curve) and second bifurcation (green curve) in the ($\mathrm{Sp}$, $\mu_a$) parameter space.} \vspace{-0.0cm}
	\label{fig:linstab}
\end{figure}


\subsection{Nonlinear dynamics, waveforms, and  trajectory analysis \label{sec:nonlinear}}

We now proceed to analyze the beating patterns and associated swimming trajectories from our nonlinear simulations, where we explore the dynamics in the ($\mathrm{Sp}$, $\mu_a$) parameter space. Consistent with the results from the linear stability analysis of Sec.~3\ref{sec:stability}, we find that spontaneous oscillations only emerge above a critical level of activity that is dependent on sperm number and coincides with the Hopf bifurcation identified in Fig.~\ref{fig:linstab}. Above the bifurcation, the flagellum starts beating spontaneously. For all cases analyzed here, we find that the oscillations take the form of traveling waves that propagate from the head towards the flagellum tip, giving rise to locomotion in the forward direction. Note that this differs from past nonlinear simulations of clamped filaments \cite{chakrabarti2019spontaneous,man2020cilia}, where both anterograde and retrograde wave propagation was observed depending on the level of activity, but is consistent with simulations of hinged filaments where only base-to-tip propagation was reported \cite{man2020cilia}. Indeed, freely swimming cells are subject to significant oscillations of the head, which makes them more similar to hinged filaments.

\begin{figure}[t]
	\centering
	\includegraphics[width=0.9\linewidth]{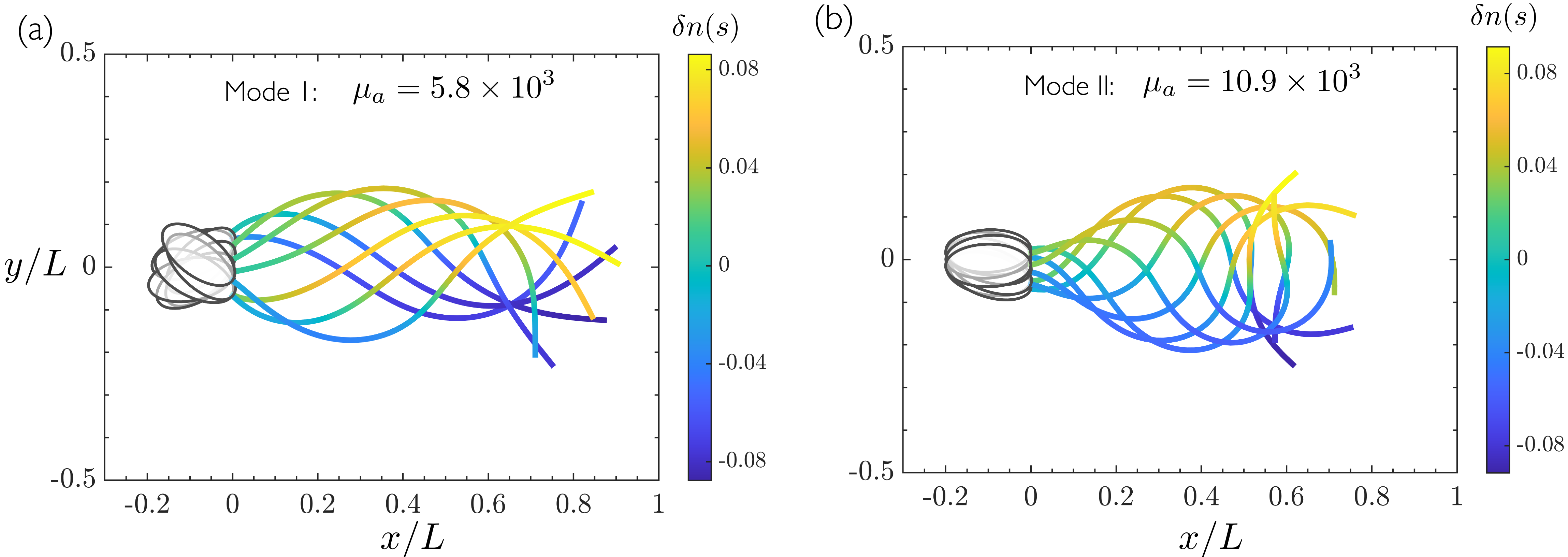}
	\caption{Superimposed snapshots of a sequence of flagellar waveforms over one period of oscillation, for two different levels of activity: (a) $\mu_a = 5820$ (mode 1) and (b) $\mu_a = 10910$ (mode 2), both for $\mathrm{Sp} = 4$. See the Supplemental Material for videos of the dynamics. }\vspace{-0.0cm}
	\label{fig:waveform}
\end{figure}

\begin{figure*}
	\centering
	\includegraphics[width=\linewidth]{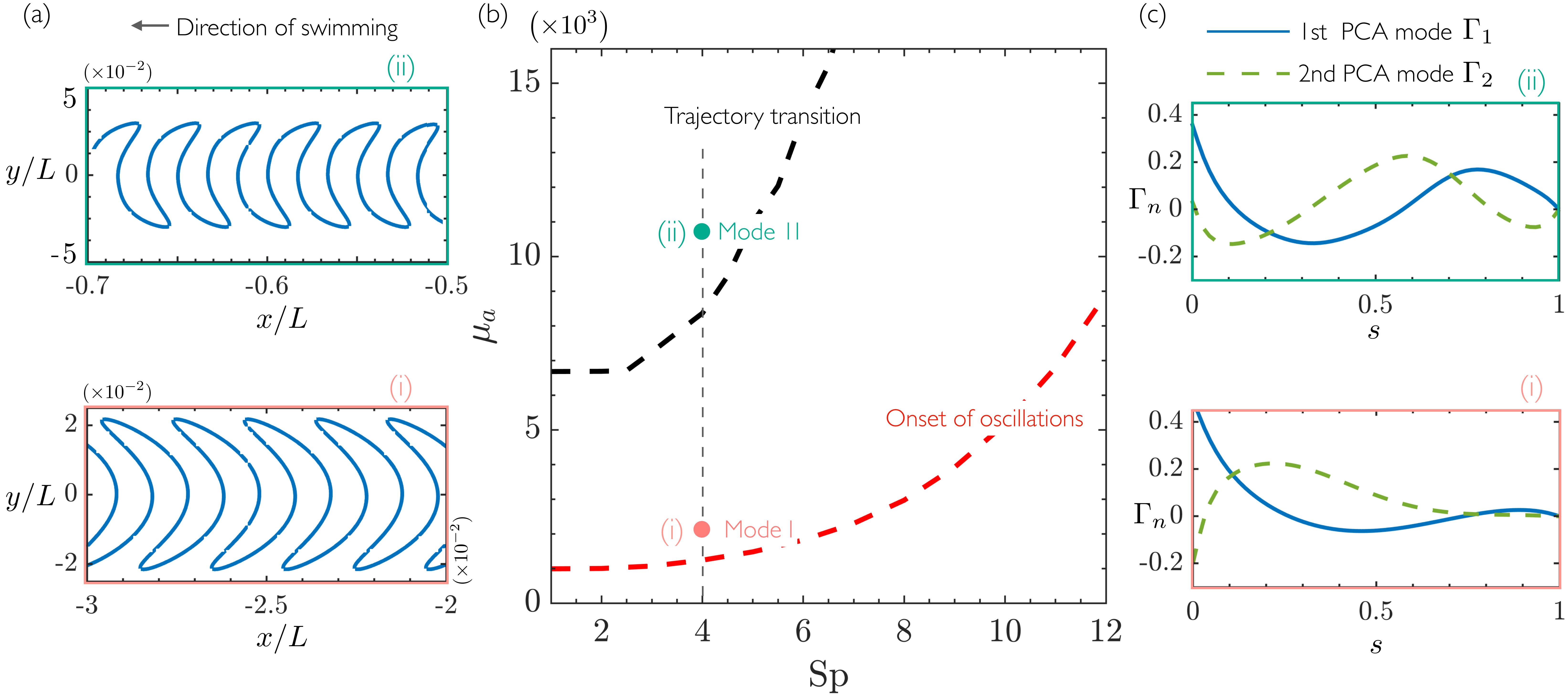}
	\caption{(a) Head trajectory for two representative cases with the same sperm number but different levels of activity: (i) $\mathrm{Sp} = 4$, $\mu_a = 1749$; (ii) $\mathrm{Sp} = 4$, $\mu_a = 11928$. The two trajectories show distinctly different behaviors, with a change in concavity which is used to identify the transition between modes 1 and 2. (b) Phase diagram in the ($\mathrm{Sp}$, $\mu_a$) parameter space highlighting the onset of spontaneous oscillations (which coincides with the Hopf bifurcation of Fig.~\ref{fig:linstab}) as well as the trajectory transition delineating modes 1 and 2. (c) Dominant PCA modes $\Gamma_1$ and $\Gamma_2$ corresponding to the same two cases as shown in (a).}
	\label{fig:PCA}
\end{figure*}

Upon exploring the ($\mathrm{Sp}$, $\mu_a$) parameter space above the Hopf bifurcation, our main finding is the existence of two qualitatively distinct swimming modes, which we proceed to characterize here. Figure~\ref{fig:waveform} overlays flagellar waveforms over one period of beating for two representative cases corresponding to the same sperm number of $\mathrm{Sp}=4$ but two distinct levels of activity $\mu_a$. Some qualitative differences between these two cases can be gleaned visually. At the lower activity level in Fig.~\ref{fig:waveform}(a), which we denote as mode 1, the waveform has a spindle-shaped envelope and involves large rotations of the sperm head with respect to the swimming direction. At the higher activity level in Fig.~\ref{fig:waveform}(b), denoted as mode 2, the waveform involves shorter-wavelength deformations and adopts a tapered shape towards the head, which displays much weaker rotations than in the previous case. 

The starkest difference between the two swimming modes is encoded in the nature of the trajectory traced by the sperm head as it swims. This is illustrated in Fig.~\ref{fig:PCA}(a), showing head trajectories for two representative cases corresponding to each swimming mode. In both cases, the trajectories involve a periodic zigzagging motion as the flagella oscillate and the cells swim from right to left. Yet, the two trajectories display opposite concavity in the $x$--$y$ plane. This clear distinction between the two types of motion allows us to systematically delineate their boundary in the ($\mathrm{Sp}$, $\mu_a$) plane as shown in Fig.~\ref{fig:PCA}(b). The trajectory transition between modes 1 and 2 roughly follows the shape of the second bifurcation identified by the linear stability analysis of Sec.~3\ref{sec:stability} but occurs at a slightly higher value of $\mu_a$, and we attribute this quantitative mismatch to nonlinearities. 

We also characterize the waveforms for both modes by applying principal component analysis (PCA) \cite{brunton2019data} on the curvature $\kappa(s,t) \equiv \partial_s \phi(s,t)$, which is decomposed as \vspace{0.1cm}
\begin{equation}
	\kappa(s,t) = \sum_n w_n(t) \Gamma_n(s), \vspace{-0.0cm}
\end{equation}
where $\Gamma_n(s)$ is the $n^{\mathrm{th}}$ PCA mode and $w_n(t)$ is the associated weight. For all the cases in our simulations, we found that the first two principal values capture more than 90\% of the spatio-temporal information \cite{chakrabarti2019hydrodynamic}. Figure~\ref{fig:PCA}(c) displays the two dominant PCA modes for the beating patterns shown in (a). As already observed in Fig.~\ref{fig:waveform}, the PCA modes reveal that larger activity results in a higher spatial frequency (shorter wavelength) in the flagellar waveform.

\begin{figure*}
	\centering
	\includegraphics[width=1\linewidth]{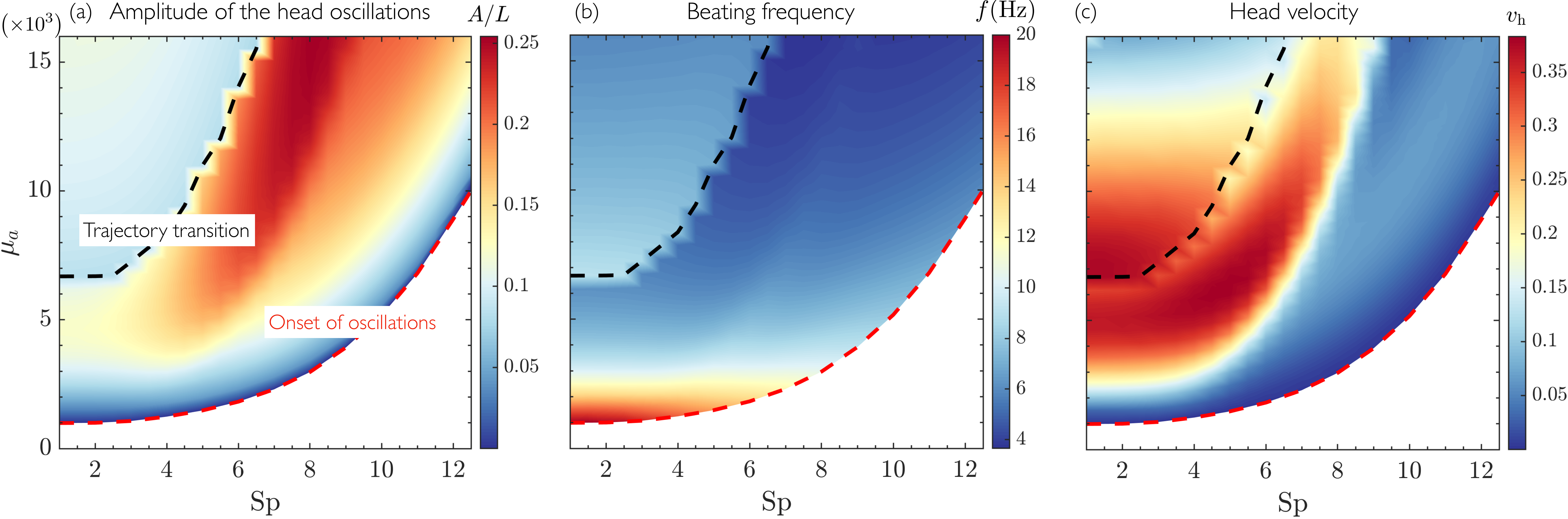}
	\caption{Variation in the ($\mathrm{Sp}$, $\mu_a$) parameter space of: (a) the amplitude $A$ of oscillation, defined as the maximum range of transverse displacement of the head over the course of one beat; (b) the beating frequency $f$, estimated in Hz for a given choice of $\tau_0=40\,\mathrm{ms}$ for the motor correlation time; and (c) the mean swimming speed $v_\text{h}$ of the cell.}
	\label{fig:afd}
\end{figure*}

We further explore the dependence of the swimming characteristics on sperm number and activity parameter in Fig.~\ref{fig:afd}, where we show the amplitude of the head oscillations, frequency of beating and swimming velocity in the ($\mathrm{Sp}$, $\mu_a$) parameter space. Very close to the Hopf bifurcation marking the onset of spontaneous oscillations, the sperm cell oscillates with a very small amplitude as seen in Fig.~\ref{fig:afd}(a). Here, the amplitude $A$ is defined as the maximum range of transverse displacement of the head over the course of one beat, i.e., the range of $y$ values in the plots of Fig.~\ref{fig:PCA}(a). As the activity $\mu_a$ increases, so does the amplitude of the head oscillations, which is largest for mode 1 at intermediate sperm numbers close to the trajectory transition. As the transition from mode 1 to mode 2 takes place, the amplitude decreases sharply as the nature of the flagellar beat changes, and only very slightly increases again upon further increasing $\mu_a$. As shown in Fig.~\ref{fig:afd}(b), the beat frequency is primarily controlled by dynein activity, and decreases monotonically with $\mu_a$ except across the trajectory transition, where it undergoes a positive jump. Similar trends had previously been reported for clamped filaments \cite{chakrabarti2019spontaneous}. The initial decrease of frequency with $\mu_a$ was previously explained by Oriola \textit{et al.} \cite{oriola2017nonlinear} as follows: a larger value of $\mu_a$ can be interpreted as a larger number density $\rho$ of dynein motors along the flagellum, resulting in an increase in the time needed for the coordinated binding and unbinding of these motors and therefore a decrease in the oscillation frequency. Setting a motor correlation time of $\tau_0 = 40 \ \mathrm{ms}$ results in dimensional frequencies in the range of $f \sim 10-20 \ \mathrm{Hz}$, consistent with that of mammalian spermatozoa \cite{gaffney2011mammalian}. We finally show the swimming velocity $v_\mathrm{h}$ in Fig.~\ref{fig:afd}(c), which is calculated as the average velocity of the sperm head in the $x$ direction over one beating period. Close to the Hopf bifurcation and onset of oscillations, the beating amplitude is very small as previously seen in Fig.~\ref{fig:afd}(a), resulting in a negligible swimming speed. With increasing activity, the swimming speed starts to increase, and displays two distinct maxima, one for each swimming mode, and both located close to the trajectory transition. Further increasing activity past the transition to mode 2 ultimately results in a decrease in the swimming speed, as the energy input is spent in side-wise swaying motion of the flagellum \cite{chakrabarti2019spontaneous} with little propulsion over one beating period. These trends will be further illustrated in our following discussion on efficiency of swimming.

\subsection{Swimming efficiency}

In order to quantify the performance of the swimming cell, we first probe into the energy budget of our active spermatozoon model. This is given by \cite{liu2018morphological}
\begin{equation}\label{eq:budget}
	\frac{\md E}{\md t} + P_{d} = P_{a},
\end{equation}
where $E$ is the net bending energy stored in the flagellum and $P_d$ is the total viscous dissipation in the suspending fluid. $P_a$ is the active power input provided by the ATP-consuming dynein motors and can be expressed as
\begin{equation}
	P_a = \int_0^1 f_m \dot{\Delta} \ \md s,
\end{equation}
where $f_m$ is the sliding force in the axoneme and $\dot{\Delta}$ is the sliding velocity. Integrating equation~\eqref{eq:budget} over one period of oscillation $T$ yields
\begin{equation}
	W_{d} = \int_{\tau}^{\tau+T} P_{d} \ \md t = \int_{\tau}^{\tau+T} P_{a} \ \md t,
\end{equation}
where we have used the fact that $E(\tau) = E(\tau+T)$ in the steady state of beating. The above relation points to the fact that, at steady state, the net active power input by the dynein motors is fully dissipated in the viscous medium over one period of oscillation. 

\begin{figure}[t]
	\centering
	\includegraphics[width=0.45\linewidth]{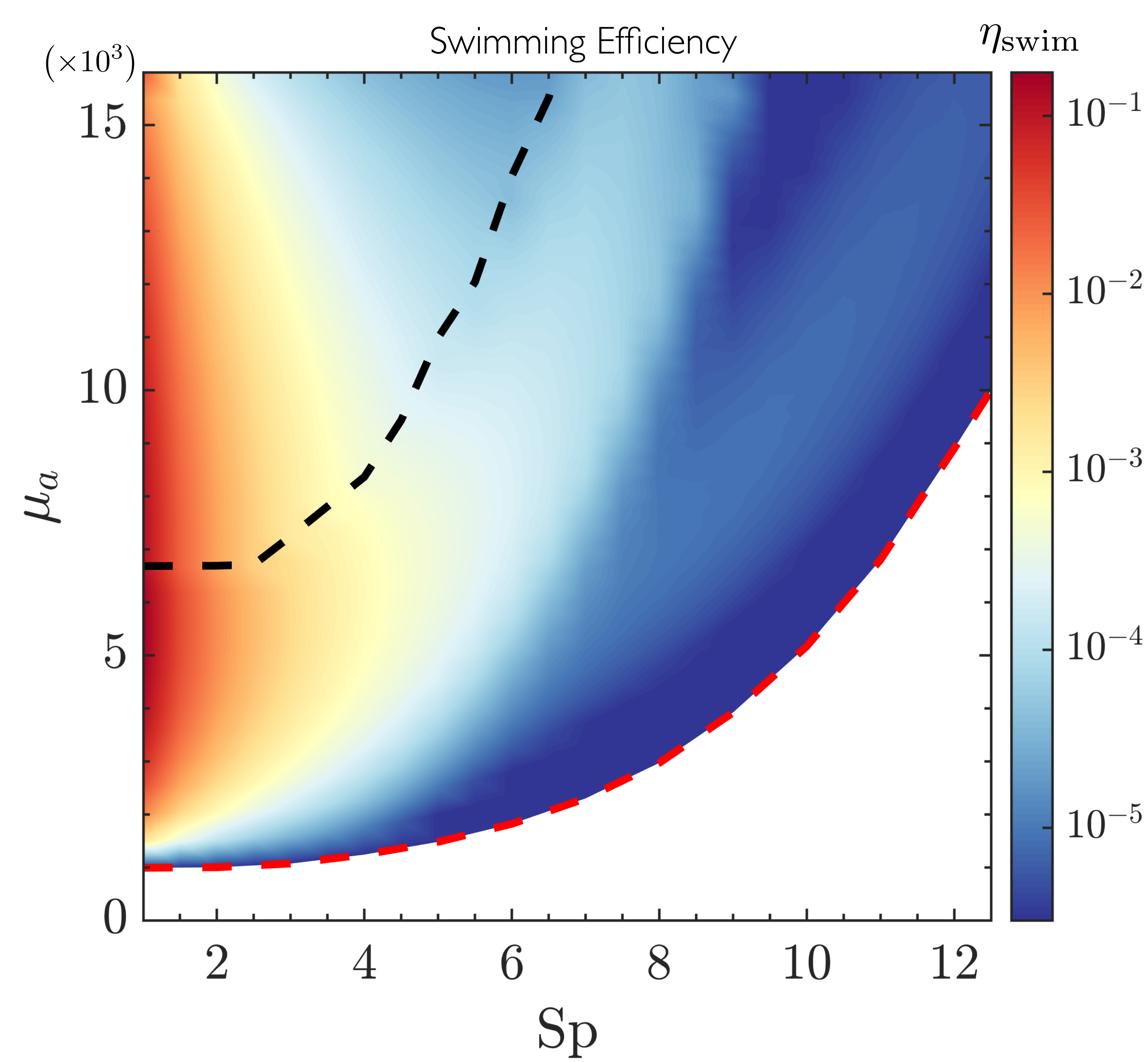}
	\caption{Variation of the swimming efficiency $\eta_\mathrm{swim}$ of the sperm cell, defined in equation~(\ref{eq:efficiency}), in the ($\mathrm{Sp}$, $\mu_a$) parameter space.}
	\label{fig:eta}
\end{figure}

As a baseline with which to compare $W_d$, we also consider an idealized sperm head translating along its major axis at a constant velocity equal to the swimming speed $v_\mathrm{h}$. The minimum energy expenditure required for the cell head to translate at that velocity for a duration of $T$ is then given by $W_\mathrm{ideal} = C_f v_\mathrm{h}^2 T$, where $C_f$ is the translational resistance coefficient of the spermatozoon head for motion along its major axis, which is known analytically for a spheroidal shape \cite{Happel1965}. This allows us to define the efficiency of the spontaneously swimming cell as
\begin{equation}
	\eta_{\mathrm{swim}} = \frac{W_\mathrm{ideal}}{W_{d}}. \label{eq:efficiency}
\end{equation}
Figure~\ref{fig:eta} shows the variation of the swimming efficiency in the ($\mathrm{Sp}$, $\mu_a$) parameter space. We find that the efficiency is primarily governed by the sperm number, and is maximum for low values of $\mathrm{Sp}$, corresponding to stiffer flagella. The swimming efficiency shows a weak dependence on $\mu_a$ and peaks at low $\mathrm{Sp}$ near the transition between the two modes of swimming, with a maximum value on the order of $10\,\%$. Comparison with Fig.~\ref{fig:afd}(c) also shows that the efficiency is positively correlated with the mean swimming speed. 


\subsection{Flow fields}

Finally, we proceed to discuss features of the flow fields generated by the swimming sperm. The dimensionless instantaneous velocity at any point $\bx$ in the fluid is obtained as the disturbance induced by the distribution of tractions along the flagellum and on the surface of the head:
\begin{equation}
	\begin{split}
		\bu(\bx,t) = \frac{\xi_\perp}{\mathrm{Sp}^4}\Biggr[& \int_0^1 \mathbf{G}(\bx,\bx_\text{f}(s,t)) \cdot \bff_\text{f}(s,t) \ \md s  \\ 
		& + \iint_{\mathcal{D}_\text{h}} \mathbf{G}(\bx,\bx_\text{h}(t)) \cdot \bff_\text{h}(t) \ \md S(\bx_\text{h})\Biggr].
	\end{split}
\end{equation}
Figure~\ref{fig:velinst}(a) shows a snapshot of the streamlines of the instantaneous velocity field in the plane of motion superimposed on top of the velocity magnitude for a typical simulation, whereas Fig.~\ref{fig:velinst}(b) shows the associated out-of-plane vorticity field; see videos of both fields in the Supplemental Material.\ Consistent with previous simulations by Ishimoto \textit{et al.} \cite{ishimoto2017coarse} that reconstruct flagellar beating patterns and flow fields from experiments, our velocity field is characterized by a pair of counter-rotating vortices straddling the beating flagellum. As the flagellum oscillates and beats periodically, these two vortical structures periodically change direction. The signature of these dynamics in the vorticity field of Fig.~\ref{fig:velinst}(b) takes the form of alternating regions of positive and negative vorticity surrounding the flagellum, which are generated at the front of the oscillating head and propagate along with the bending wave towards the distal end of the flagellum where they vanish. 

\begin{figure}[t]
	\centering
	\includegraphics[width=0.85\linewidth]{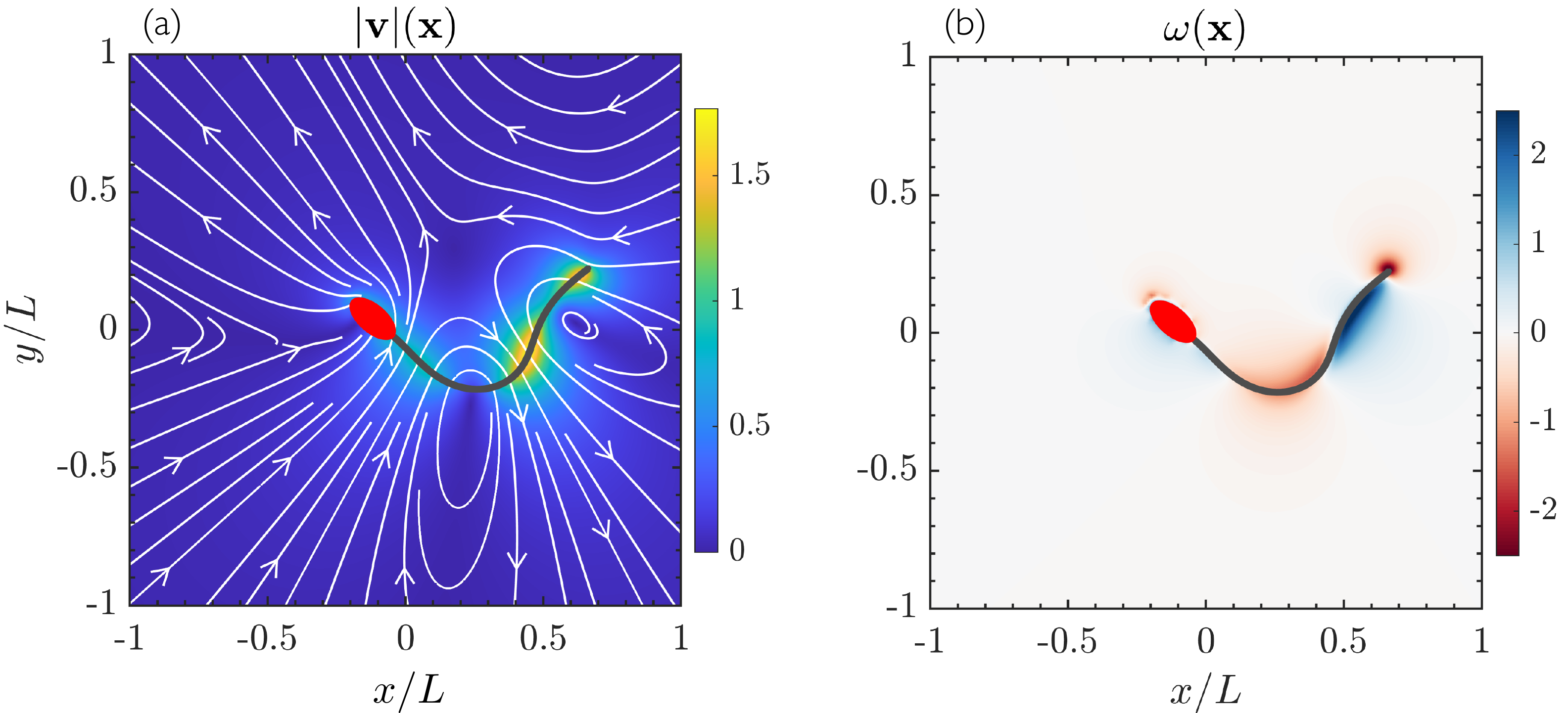}
	\caption{(a) Snapshot of the instantaneous velocity field induced by a swimming sperm in the plane of motion. The plot shows streamlines of the flow superimposed on a map of the velocity magnitude. (b) Corresponding out-of-plane vorticity field. Parameter values: $\mathrm{Sp} = 5$, $\mu_a = 8000$. See the Supplemental Material for videos of the velocity and vorticity fields.}
	\label{fig:velinst}
\end{figure}

The time-averaged flow field for the same simulation is shown in Fig.~\ref{fig:velav}(a). In the plane of motion, it is characterized in the near field by two pairs of counter-rotating vortices, one surrounding the sperm head, and the other on both sides of the flagellum, where the largest mean velocity is also observed. In the far field, the symmetry of the average flow is that of an extensile Stokes dipole, characteristic of a pusher, in which the fluid is pushed forward ahead of the cell by the moving head and expelled backward in the rear of the flagellum \cite{lauga2009hydrodynamics}. The features of this time-averaged velocity field are once again very similar to those observed in the reconstructed flow fields of Ishimoto \textit{et al.} \cite{ishimoto2017coarse}. The dipolar nature of the flow field is confirmed in Fig.~\ref{fig:velav}(b), showing the decay of the velocity magnitude in the axial and transverse directions, where a clear $1/r^2$ dependence is observed in the far field. This is in contrast to the previously analyzed velocity fields of clamped active filaments \cite{chakrabarti2019spontaneous} that showed a $1/r$ decay in the velocity field.

\begin{figure}[t]
	\centering
	\includegraphics[width=0.85\linewidth]{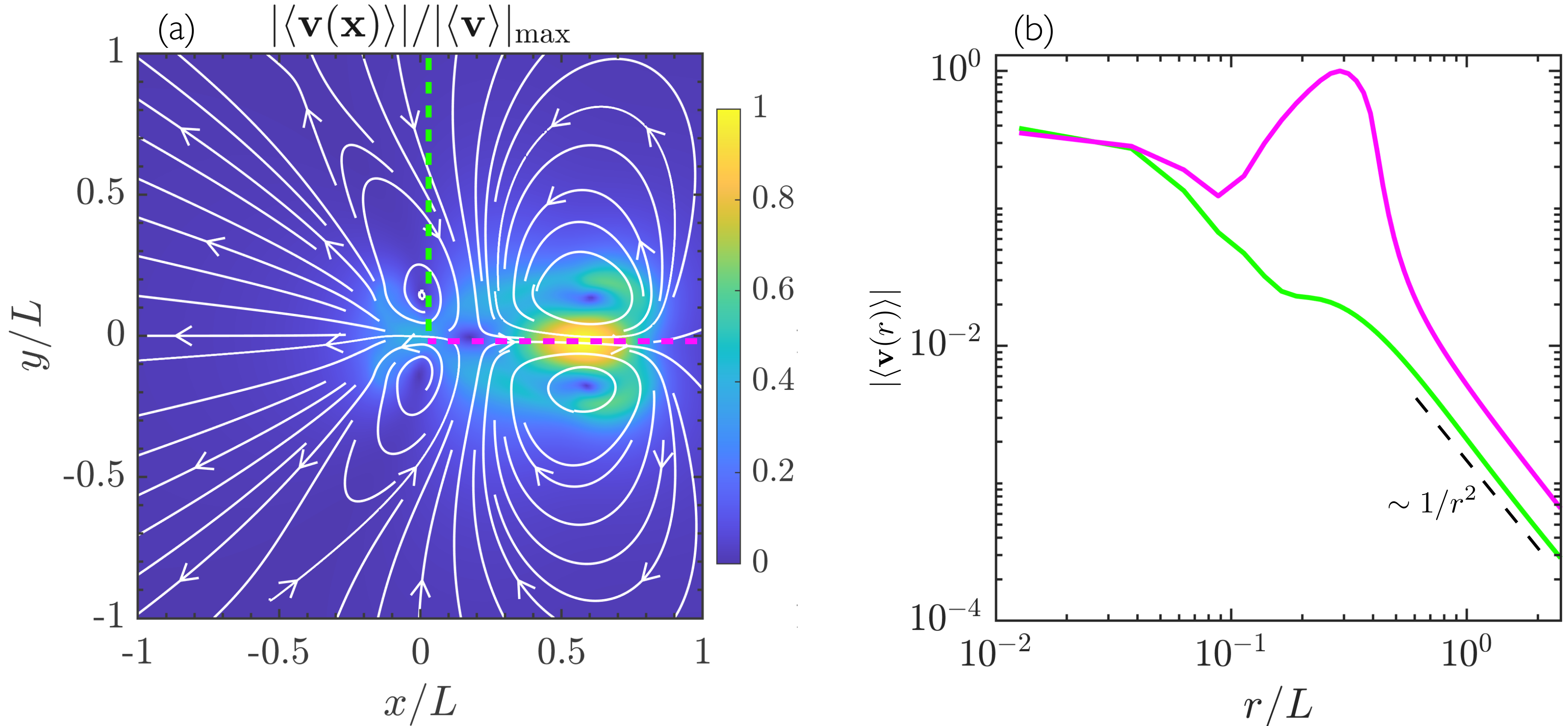}
	\caption{(a) Time-averaged velocity field for the same simulation as in Fig.~\ref{fig:velinst}, where the velocity magnitude has been normalized by its maximum value. (b) Dependence of the velocity magnitude on distance from the flagellum tip along the positive $x$ and $y$ directions marked as dotted lines in (a).}
	\label{fig:velav}
\end{figure}
\section{Conclusion\label{sec:concl}}

We have used a sliding-control active filament model to study the propulsion and hydrodynamics of an idealized swimming spermatozoon. Our model captures a feedback loop between the internal actuation by the ATP-powered dynein motors and the elastohydrodynamics of the deforming flagellum and accounts for hydrodynamic interactions. Our simulations revealed that, following a Hopf bifurcation, the flagellum starts to beat spontaneously with bending waves always propagating from the head towards the tail. This anterograde wave propagation is typical in swimming sperm cells but was absent for small activities in the previously studied case of clamped active filaments \cite{chakrabarti2019spontaneous}. Flagellar deformations are accompanied by oscillations of the cell head, modeled as a rigid spheroid, and result in locomotion in the forward direction. Flow fields computed from our simulations correlate well with those observed in experiments \cite{gaffney2011mammalian} and display a characteristic $1/r^2$ decay in the far field, confirming that swimming spermatozoa behave as force dipoles \cite{lauga2009hydrodynamics}.

Exploration of the ($\mathrm{Sp}$, $\mu_a$) parameter space highlighted a transition at a critical value of $\mu_a$ between two distinct modes of swimming, with qualitatively different kinematics and waveforms. Specifically, we showed that increasing activity above the transition results in a reversal of the concavity of the head trajectory, and in the emergence of high-wavenumber deformation modes in the flagellar waveform. An analysis of the mean swimming velocity demonstrated that it does not vary monotonically with motor activity, but instead displays two maxima, each associated with one of the two swimming modes. The prediction by our simulations of the existence of these two modes suggests an important role for the regulation of dynein activity in the sperm flagellum. It is known that an increase in dynein activity, triggered by increased levels of calcium in the flagellum \cite{BYTS2000,S2002}, is associated with sperm hyperactivation before fertilization \cite{ho2001hyperactivation}, which is accompanied by a qualitative change in the flagellar waveform and cell kinematics. The precise relationship between the two swimming modes identified by our model and the transition to hyperactivation remains, however, to be established. 

While our model and simulations capture many salient features of sperm locomotion, they rely on several simplifying assumptions that could be relaxed in future work in order to allow for more direct comparison with experiments. In particular, we have assumed a simple shape for the sperm head and uniform radius and mechanical properties for the flagellum: in reality, sperm morphology can vary significantly among species as well as among males across a population \cite{PHB2009}. Incorporating more detailed morphological and mechanical features may be essential to capture certain characteristics of sperm locomotion \cite{GG2019} and may also help shed light on variations in sperm performance \cite{HES2008}. Our current model is also limited to planar deformations, while recent experimental evidence suggests that flagellar deformations may in fact be three-dimensional \cite{GHMDC2020,PPNGPOPSN2022}: accounting for 3D deformations would require a more detailed description of the axonemal structure and of the elastodynamics of the flagellum, for instance as an active Kirkhoff rod \cite{Antman:1250280}. Other improvements to the present work could include a model for internal dissipation inside the flagellum, which may be significant as suggested by recent observations \cite{NGPNSOJP2021}, as well as for the coupling of dynein activity with calcium signaling \cite{olson2011coupling,olson2010model,OFS2011}. Finally, we note that our model is well-suited to study the phenomenon of hydrodynamic synchronization between swimming spermatozoa \cite{yang2008cooperation,WIG2019}, in the spirit of our past work on clamped flagella \cite{chakrabarti2019hydrodynamic}. This synchronization phenomenon is key to various collective dynamics in suspensions of sperm cells \cite{CPDKP2015}, which up to now have been primarily studied using models that prescribe flagellar kinematics \cite{schoeller2018flagellar,SHK2020}.

\bibliographystyle{ieeetr}
\bibliography{spermrefs}

\providecommand{\noopsort}[1]{}\providecommand{\singleletter}[1]{#1}%
\begin{thebibliography}{10}

\bibitem{purcell1977life}
E.~M. Purcell, ``Life at low reynolds number,'' {\em Am. J. Phys.}, vol.~45,
  no.~1, pp.~3--11, 1977.

\bibitem{lauga2020fluid}
E.~Lauga, {\em The Fluid Dynamics of Cell Motility}.
\newblock Cambridge University Press, 2020.

\bibitem{lauga2009hydrodynamics}
E.~Lauga and T.~R. Powers, ``The hydrodynamics of swimming microorganisms,''
  {\em Rep. Prog. Phys.}, vol.~72, no.~9, p.~096601, 2009.

\bibitem{alberts2015essential}
B.~Alberts, D.~Bray, K.~Hopkin, A.~D. Johnson, J.~Lewis, M.~Raff, K.~Roberts,
  and P.~Walter, {\em Essential Cell Biology}.
\newblock Garland Science, 2015.

\bibitem{brokaw1972computer}
C.~J. Brokaw, ``Computer simulation of flagellar movement: I. demonstration of
  stable bend propagation and bend initiation by the sliding filament model,''
  {\em Biophys. J.}, vol.~12, no.~5, pp.~564--586, 1972.

\bibitem{riedel2007molecular}
I.~H. Riedel-Kruse, A.~Hilfinger, J.~Howard, and F.~J{\"u}licher, ``How
  molecular motors shape the flagellar beat,'' {\em HFSP J.}, vol.~1, no.~3,
  pp.~192--208, 2007.

\bibitem{chakrabarti2019spontaneous}
B.~Chakrabarti and D.~Saintillan, ``Spontaneous oscillations, beating patterns,
  and hydrodynamics of active microfilaments,'' {\em Phys. Rev. Fluids},
  vol.~4, no.~4, p.~043102, 2019.

\bibitem{brokaw1972flagellar}
C.~J. Brokaw, ``Flagellar movement: a sliding filament model: an explanation is
  suggested for the spontaneous propagation of bending waves by flagella,''
  {\em Science}, vol.~178, no.~4060, pp.~455--462, 1972.

\bibitem{brokaw1999computer}
C.~J. Brokaw, ``Computer simulation of flagellar movement: Vii. conventional
  but functionally different cross-bridge models for inner and outer arm
  dyneins can explain the effects of outer arm dynein removal,'' {\em Cell
  Motil. Cytoskelet.}, vol.~42, no.~2, pp.~134--148, 1999.

\bibitem{brokaw2002computer}
C.~J. Brokaw, ``Computer simulation of flagellar movement viii: coordination of
  dynein by local curvature control can generate helical bending waves,'' {\em
  Cell Motil. Cytoskelet.}, vol.~53, no.~2, pp.~103--124, 2002.

\bibitem{brokaw2005computer}
C.~J. Brokaw, ``Computer simulation of flagellar movement ix. oscillation and
  symmetry breaking in a model for short flagella and nodal cilia,'' {\em Cell
  Motil. Cytoskelet.}, vol.~60, no.~1, pp.~35--47, 2005.

\bibitem{brokaw2014computer}
C.~J. Brokaw, ``Computer simulation of flagellar movement x: doublet pair
  splitting and bend propagation modeled using stochastic dynein kinetics,''
  {\em Cytoskeleton}, vol.~71, no.~4, pp.~273--284, 2014.

\bibitem{lindemann1994geometric}
C.~B. Lindemann, ``A ``geometric clutch'' hypothesis to explain oscillations of
  the axoneme of cilia and flagella,'' {\em J. Theor. Biol.}, vol.~168, no.~2,
  pp.~175--189, 1994.

\bibitem{lindemann1994model}
C.~B. Lindemann, ``A model of flagellar and ciliary functioning which uses the
  forces transverse to the axoneme as the regulator of dynein activation,''
  {\em Cell Motil. Cytoskelet.}, vol.~29, no.~2, pp.~141--154, 1994.

\bibitem{lindemann1996functional}
C.~B. Lindemann, ``Functional significance of the outer dense fibers of
  mammalian sperm examined by computer simulations with the geometric clutch
  model,'' {\em Cell Motil. Cytoskelet.}, vol.~34, no.~4, pp.~258--270, 1996.

\bibitem{holcomb1999flagellar}
D.~L. Holcomb-Wygle, K.~A. Schmitz, and C.~B. Lindemann, ``Flagellar arrest
  behavior predicted by the geometric clutch model is confirmed experimentally
  by micromanipulation experiments on reactivated bull sperm,'' {\em Cell
  Motil. Cytoskelet.}, vol.~44, no.~3, pp.~177--189, 1999.

\bibitem{lindemann2002geometric}
C.~B. Lindemann, ``Geometric clutch model version 3: The role of the inner and
  outer arm dyneins in the ciliary beat,'' {\em Cell Motil. Cytoskelet.},
  vol.~52, no.~4, pp.~242--254, 2002.

\bibitem{bayly2014equations}
P.~V. Bayly and K.~S. Wilson, ``Equations of interdoublet separation during
  flagella motion reveal mechanisms of wave propagation and instability,'' {\em
  Biophys. J.}, vol.~107, no.~7, pp.~1756--1772, 2014.

\bibitem{oriola2017nonlinear}
D.~Oriola, H.~Gad{\^e}lha, and J.~Casademunt, ``Nonlinear amplitude dynamics in
  flagellar beating,'' {\em Royal Soc. Open Sci.}, vol.~4, no.~3, p.~160698,
  2017.

\bibitem{chakrabarti2019hydrodynamic}
B.~Chakrabarti and D.~Saintillan, ``Hydrodynamic synchronization of
  spontaneously beating filaments,'' {\em Phys. Rev. Lett.}, vol.~123, no.~20,
  p.~208101, 2019.

\bibitem{taylor1951analysis}
G.~I. Taylor, ``Analysis of the swimming of microscopic organisms,'' {\em Proc.
  R. Soc. A: Math. Phys. Eng. Sci.}, vol.~209, no.~1099, pp.~447--461, 1951.

\bibitem{taylor1952analysis}
G.~I. Taylor, ``Analysis of the swimming of long and narrow animals,'' {\em
  Proc. R. Soc. A: Math. Phys. Eng. Sci.}, vol.~214, no.~1117, pp.~158--183,
  1952.

\bibitem{hancock1953self}
G.~Hancock, ``The self-propulsion of microscopic organisms through liquids,''
  {\em Proc. R. Soc. A: Math. Phys. Eng. Sci.}, vol.~217, no.~1128,
  pp.~96--121, 1953.

\bibitem{gray1955propulsion}
J.~Gray and G.~Hancock, ``The propulsion of sea-urchin spermatozoa,'' {\em J.
  Exp. Biol.}, vol.~32, no.~4, pp.~802--814, 1955.

\bibitem{dresdner1980propulsion}
R.~Dresdner, D.~Katz, and S.~Berger, ``The propulsion by large amplitude waves
  of uniflagellar micro-organisms of finite length,'' {\em J. Fluid Mech.},
  vol.~97, no.~3, pp.~591--621, 1980.

\bibitem{higdon1979hydrodynamic}
J.~J. Higdon, ``A hydrodynamic analysis of flagellar propulsion,'' {\em J.
  Fluid Mech.}, vol.~90, no.~4, pp.~685--711, 1979.

\bibitem{phan1987boundary}
N.~Phan-Thien, T.~Tran-Cong, and M.~Ramia, ``A boundary-element analysis of
  flagellar propulsion,'' {\em J. Fluid Mech.}, vol.~184, pp.~533--549, 1987.

\bibitem{ramia1993role}
M.~Ramia, D.~Tullock, and N.~Phan-Thien, ``The role of hydrodynamic interaction
  in the locomotion of microorganisms,'' {\em Biophys. J.}, vol.~65, no.~2,
  pp.~755--778, 1993.

\bibitem{fauci1995sperm}
L.~J. Fauci and A.~McDonald, ``Sperm motility in the presence of boundaries,''
  {\em Bull. Math. Biol.}, vol.~57, no.~5, pp.~679--699, 1995.

\bibitem{elgeti2010hydrodynamics}
J.~Elgeti, U.~B. Kaupp, and G.~Gompper, ``Hydrodynamics of sperm cells near
  surfaces,'' {\em Biophys. J.}, vol.~99, no.~4, pp.~1018--1026, 2010.

\bibitem{smith2009human}
D.~Smith, E.~Gaffney, J.~Blake, and J.~Kirkman-Brown, ``Human sperm
  accumulation near surfaces: a simulation study,'' {\em J. Fluid Mech.},
  vol.~621, pp.~289--320, 2009.

\bibitem{teran2010viscoelastic}
J.~Teran, L.~Fauci, and M.~Shelley, ``Viscoelastic fluid response can increase
  the speed and efficiency of a free swimmer,'' {\em Phys. Rev. Lett.},
  vol.~104, no.~3, p.~038101, 2010.

\bibitem{lauga2007propulsion}
E.~Lauga, ``Propulsion in a viscoelastic fluid,'' {\em Phys. Fluids}, vol.~19,
  no.~8, p.~083104, 2007.

\bibitem{gillies2009hydrodynamic}
E.~A. Gillies, R.~M. Cannon, R.~B. Green, and A.~A. Pacey, ``Hydrodynamic
  propulsion of human sperm,'' {\em J. Fluid Mech.}, vol.~625, pp.~445--474,
  2009.

\bibitem{gadelha2010nonlinear}
H.~Gad{\^e}lha, E.~Gaffney, D.~Smith, and J.~Kirkman-Brown, ``Nonlinear
  instability in flagellar dynamics: a novel modulation mechanism in sperm
  migration?,'' {\em J. R. Soc. Interface}, vol.~7, no.~53, pp.~1689--1697,
  2010.

\bibitem{dillon2003mathematical}
R.~Dillon, L.~Fauci, and C.~Omoto, ``Mathematical modeling of axoneme mechanics
  and fluid dynamics in ciliary and sperm motility,'' {\em Dyn. Contin.
  Discrete Impuls. Syst. A: Math. Anal.}, vol.~10, pp.~745--758, 2003.

\bibitem{olson2011coupling}
S.~D. Olson, S.~S. Suarez, and L.~J. Fauci, ``Coupling biochemistry and
  hydrodynamics captures hyper-activated sperm motility in a simple flagellar
  model,'' {\em J. Theor. Biol.}, vol.~283, no.~1, pp.~203--216, 2011.

\bibitem{ho2001hyperactivation}
H.-C. Ho and S.~S. Suarez, ``Hyperactivation of mammalian spermatozoa: function
  and regulation,'' {\em Reproduction}, vol.~122, no.~4, pp.~519--526, 2001.

\bibitem{olson2010model}
S.~D. Olson, S.~S. Suarez, and L.~J. Fauci, ``A model of catsper channel
  mediated calcium dynamics in mammalian spermatozoa,'' {\em Bull. Math.
  Biol.}, vol.~72, no.~8, pp.~1925--1946, 2010.

\bibitem{gaffney2011mammalian}
E.~A. Gaffney, H.~Gad{\^e}lha, D.~J. Smith, J.~R. Blake, and J.~C.
  Kirkman-Brown, ``Mammalian sperm motility: observation and theory,'' {\em
  Annu. Rev. Fluid Mech.}, vol.~43, pp.~501--528, 2011.

\bibitem{Antman:1250280}
S.~Antman, {\em {Nonlinear Problems of Elasticity}}.
\newblock Springer, 2005.

\bibitem{KR1976}
J.~B. Keller and S.~I. Rubinow, ``Slender-body theory for slow viscous flow,''
  {\em J. Fluid Mech.}, vol.~75, pp.~705--714, 1976.

\bibitem{tornberg2004simulating}
A.-K. Tornberg and M.~J. Shelley, ``Simulating the dynamics and interactions of
  flexible fibers in stokes flows,'' {\em J. Comput. Phys.}, vol.~196, no.~1,
  pp.~8--40, 2004.

\bibitem{pozrikidis1992}
C.~Pozrikidis, {\em Boundary Integral and Singularity Methods for Linearized
  Viscous Flow}.
\newblock Cambridge University Press, 1992.

\bibitem{sartori2019effect}
P.~Sartori, {\em Effect of curvature and normal forces on motor regulation of
  cilia}.
\newblock PhD thesis, Technische {Universit\"at} Dresden, 2019.

\bibitem{pozrikidis2002practical}
C.~Pozrikidis, {\em A Practical Guide to Boundary Element Methods with the
  Software Library BEMLIB}.
\newblock CRC Press, 2002.

\bibitem{howard2001mechanics}
J.~Howard, {\em Mechanics of Motor Proteins and the Cytoskeleton}.
\newblock Sunderland, MA: Sinauer, 2001.

\bibitem{man2020cilia}
Y.~Man, F.~Ling, and E.~Kanso, ``Cilia oscillations,'' {\em Philos. Trans. R.
  Soc. B}, vol.~375, no.~1792, p.~20190157, 2020.

\bibitem{brunton2019data}
J.~N. Kutz, {\em Data-Driven Modeling \& Scientific Computation: Methods for
  Complex Systems \& Big Data}.
\newblock Oxford University Press, 2013.

\bibitem{liu2018morphological}
Y.~Liu, B.~Chakrabarti, D.~Saintillan, A.~Lindner, and O.~du~Roure,
  ``Morphological transitions of elastic filaments in shear flow,'' {\em Proc.
  Natl. Acad. Sci. USA}, vol.~115, no.~38, pp.~9438--9443, 2018.

\bibitem{Happel1965}
J.~Happel and H.~Brenner, {\em {Low Reynolds Number Hydrodynamics}}.
\newblock Prentice-Hall, 1965.

\bibitem{ishimoto2017coarse}
K.~Ishimoto, H.~Gad{\^e}lha, E.~A. Gaffney, D.~J. Smith, and J.~Kirkman-Brown,
  ``Coarse-graining the fluid flow around a human sperm,'' {\em Phys. Rev.
  Lett.}, vol.~118, no.~12, p.~124501, 2017.

\bibitem{BYTS2000}
H.~Bannai, M.~Yoshimure, K.~Takahashi, and C.~Shingyoji, ``Calcium regulation
  of microtubule sliding in reactivated sea urchin sperm flagella,'' {\em J.
  Cell Sci.}, vol.~113, pp.~831--839, 2000.

\bibitem{S2002}
E.~F. Smith, ``Regulation of flagellar dynein by calcium and a role for an
  axonemal calmodulin and calmodulin-dependent kinase,'' {\em Mol. Biol. Cell},
  vol.~13, pp.~3303--3313, 2002.

\bibitem{PHB2009}
S.~Pitnick, D.~J. Hosken, and T.~R. Birkhead, ``Sperm morphological
  diversity,'' in {\em Sperm Biology: An Evolutionary Perspective} (T.~R.
  Birkhead, D.~J. Hosken, and S.~Pitnick, eds.), ch.~3, pp.~69--149, Oxford,
  U.K.: Academic Press, 2009.

\bibitem{GG2019}
H.~{Gad\^elha} and E.~A. Gaffney, ``Flagellar ultrastructure suppresses
  buckling instabilities and enables mammalian sperm navigation in
  high-viscosity media,'' {\em J. R. Soc. Interface}, vol.~16, p.~20180668,
  2019.

\bibitem{HES2008}
S.~Humphries, J.~P. Evans, and L.~W. Simmons, ``Sperm competition: linking form
  to function,'' {\em BMC Evol. Biol.}, vol.~8, p.~319, 2008.

\bibitem{GHMDC2020}
H.~{Gad\^elha}, P.~{Hern\'andez-Herrera}, F.~Montoya, A.~Darszon, and
  G.~Corkidi, ``Human sperm uses asymmetric and anisotropic flagellar controls
  to regulate swimming symmetry and cell steering,'' {\em Sci. Adv.}, vol.~6,
  p.~eaba5168, 2020.

\bibitem{PPNGPOPSN2022}
S.~Powar, F.~Y. Parast, A.~Nandagiri, A.~S. Gaikwad, D.~L. Potter, M.~K.
  {O'Bryan}, R.~Prabhakar, J.~Soria, and R.~Nosrati, ``Unraveling the
  kinematics of sperm motion by reconstructing the flagellar wave motion in
  3d,'' {\em Small Methods}, p.~2101089, 2022.

\bibitem{NGPNSOJP2021}
A.~Nandagiri, A.~S. Gaikwad, D.~L. Potter, R.~Nosrati, J.~Soria, M.~K.
  {O'Bryan}, S.~Jadhav, and R.~Prabhakar, ``Flagellar energetics from
  high-resolution imaging of beating patterns in tethered mouse sperm,'' {\em
  eLife}, vol.~10, p.~e62524, 2021.

\bibitem{OFS2011}
S.~D. Olson, L.~J. Fauci, and S.~S. Suarez, ``Mathematic modeling of calcium
  signaling during sperm hyperactivation,'' {\em Mol. Hum. Reprod.}, vol.~17,
  pp.~500--510, 2011.

\bibitem{yang2008cooperation}
Y.~Yang, J.~Elgeti, and G.~Gompper, ``Cooperation of sperm in two dimensions:
  synchronization, attraction, and aggregation through hydrodynamic
  interactions,'' {\em Phys. Rev. E}, vol.~78, no.~6, p.~061903, 2008.

\bibitem{WIG2019}
B.~J. Walker, K.~Ishimoto, and E.~A. Gaffney, ``Pairwise hydrodynamic
  interactions of synchronized spermatozoa,'' {\em Phys. Rev. Fluids}, vol.~4,
  p.~093101, 2019.

\bibitem{CPDKP2015}
A.~Creppy, O.~Praud, X.~Druart, P.~L. Kohnke, and F.~{Plourabou\'e},
  ``Turbulence of swarming sperm,'' {\em Phys. Rev. E}, vol.~92, p.~032772,
  2015.

\bibitem{schoeller2018flagellar}
S.~F. Schoeller and E.~E. Keaveny, ``From flagellar undulations to collective
  motion: predicting the dynamics of sperm suspensions,'' {\em J. R. Soc.
  Interface}, vol.~15, no.~140, p.~20170834, 2018.

\bibitem{SHK2020}
S.~F. Schoeller, W.~V. Holt, and E.~E. Keaveny, ``Collective dynamics of sperm
  cells,'' {\em Phil. Trans. R. Soc. B}, vol.~375, p.~20190384, 2020.

\end{thebibliography}

\end{document}